\documentclass[12pt,a4paper]{amsart}
\usepackage[applemac]{inputenc}
\usepackage[T1]{fontenc}
\usepackage{amssymb}
\usepackage{amsmath}
\usepackage{amsthm}
\usepackage{amscd}
\usepackage{varioref} \usepackage{paralist} 
\setdefaultenum{\upshape(i)}{}{}{}
\usepackage[pdftex]{graphicx}
\usepackage{color} 
\usepackage[pdftex,colorlinks,linkcolor=black,filecolor=black,citecolor=black,urlcolor=black]{hyperref}  

\allowdisplaybreaks

\numberwithin{equation}{section}

\newtheoremstyle{mythmstyle}%
{1.5\baselineskip}
{\baselineskip}
{\itshape}
{}
{\bf}
{}
{0pt}
{} 

\newtheoremstyle{mydefstyle}%
{1.5\baselineskip}
{\baselineskip}
{}
{}
{\bf}
{}
{0pt}
{} 

\newtheoremstyle{mypreuvestyle}%
{\baselineskip}
{\baselineskip}
{}
{}
{\em}
{}
{0pt}
{} 

\newif\ifmynonumberenvi\mynonumberenvitrue
\theoremstyle{mythmstyle}
\newtheorem{proclaimmythm}[equation]{} \newtheorem*{proclaimmythm*}{}
\newenvironment{proclaim}[2][*]{\ifx*#1\mynonumberenvitrue\begin{proclaimmythm*}{\bf#2.} \ignorespaces\else\mynonumberenvifalse\begin{proclaimmythm}{.\kern0.5em\bf#2.}\label{#1} \ignorespaces\fi}{\ifmynonumberenvi\end{proclaimmythm*}\else\end{proclaimmythm}\fi}
\renewenvironment{proclaim}[2][*]{\ifx*#1\begin{proclaimmythm}{.\kern0.5em\bf#2.} \ignorespaces\else\begin{proclaimmythm}{.\kern0.5em\bf#2.}\label{#1} \ignorespaces\fi}{\end{proclaimmythm}}

\theoremstyle{mydefstyle}
\newtheorem{proclaimmydef}[equation]{} 
\newtheorem*{proclaimmydef*}{}
\newenvironment{definition}[2][*]{\ifx*#1\mynonumberenvitrue\begin{proclaimmydef*}{\bf#2.}\else\mynonumberenvifalse\begin{proclaimmydef}{.\kern0.5em\bf#2.}\label{#1} \ignorespaces\fi}{\ifmynonumberenvi\end{proclaimmydef*}\else\end{proclaimmydef}\fi}
\renewenvironment{definition}[2][*]{\ifx*#1\begin{proclaimmydef}{.\kern0.5em\bf#2.} \ignorespaces\else\begin{proclaimmydef}{.\kern0.5em\bf#2.}\label{#1} \ignorespaces\fi}{\end{proclaimmydef}}

\def\QEDbox{\hbox{\lower2.3pt\vbox{\hrule\hbox
   {\vrule\kern1pt\vbox{\kern1.7pt\hbox{$\scriptstyle
   QED$}\kern.6pt}\kern1pt\vrule}\hrule}}}
\def\QED{\hskip0.01em plus 40pt\null{} \null\nobreak\hfill
   \kern3pt\QEDbox} 
\newcommand\QEDici{\\\noalign{\vskip-\baselineskip\smash{\hbox to\linewidth{\vrule width0pt \hfill\global\QEDdejaplacetrue\QEDbox}}\vskip-\baselineskip}}
\newif\ifQEDdejaplace\QEDdejaplacefalse

\theoremstyle{mypreuvestyle}
\newtheorem*{proclaimmypreuve}{}
\newenvironment{preuve}[1][*]{\begin{proclaimmypreuve}{\ifx*#1{\em Proof.}\else{\em#1.}\fi} \ignorespaces}{\ifQEDdejaplace\global\QEDdejaplacefalse\else\QED\fi\end{proclaimmypreuve}}

\newif\ifmynonumberequation\mynonumberequationtrue

\makeatletter
\numberwithin{equation}{section}
\newenvironment{moneq}[1][*]{\ifx*#1\mynonumberequationtrue\begin{equation*}\else\mynonumberequationfalse\begin{equation}\label{#1}\fi}{\ifmynonumberequation\end{equation*}\@ignoretrue\else\end{equation}\@ignoretrue\fi\ignorespaces}
\makeatother

\newcommand{\recalf}[1]{(\ref{#1})}
\newcommand{\recalt}[1]{{[\ref{#1}]}}

\def\dcap_#1{\mathchoice{%
          {\textstyle\bigcap\limits_{#1}}}%
          {\underset{#1}\cap}%
          {\underset{#1}\cap}%
          {\underset{#1}\cap}}
\def\dcup_#1{\mathchoice{%
          {\textstyle\bigcup\limits_{#1}}}%
          {\underset{#1}\cup}%
          {\underset{#1}\cup}%
          {\underset{#1}\cup}}
\def\ddcap_#1^#2{\mathchoice{%
          {\textstyle\bigcap\limits_{#1}^{#2}}}%
          {\overset{#2}{\underset{#1}\cap}}%
          {\overset{#2}{\underset{#1}\cap}}%
          {\overset{#2}{\underset{#1}\cap}}}
\def\ddcup_#1^#2{\mathchoice{%
          {\textstyle\bigcup\limits_{#1}^{#2}}}%
          {\overset{#2}{\underset{#1}\cup}}%
          {\overset{#2}{\underset{#1}\cup}}%
          {\overset{#2}{\underset{#1}\cup}}}

\def\oversetalign#1\to#2{\mathbin{\smash{\overset{{\text{\rlap{\hss#1}}}}{#2}}}}
\def\oversettext#1\to#2{\mathbin{\smash{\overset{{\text{{#1}}}}{#2}}}}

\def\migl({{\raise-0.2ex\hbox{\textup{\large(}}}}
\def\migr){{\raise-0.2ex\hbox{\textup{\large)}}}}
\def\Migl({{\raise-0.2ex\hbox{\textup{\Large(}}}}
\def\Migr){{\raise-0.2ex\hbox{\textup{\Large)}}}}

\newcommand\bigrestricted{{\kern1pt\vrule height3.3ex depth1.7ex width0.6pt\kern1pt}}
\newcommand\caprestricted{{\kern1pt\vrule height1.7ex depth0.5ex width0.6pt\kern1pt}}
\newcommand\midrestricted{{\kern1pt\vrule height1.7ex depth0.9ex width0.6pt\kern1pt}}

\newcommand\restricted{{\kern1pt\vrule height1.3ex depth0.5ex width0.6pt\kern1pt}}

\newcommand\AB{_{\alpha\beta}}
\newcommand\alphat{\widetilde{\alpha}}

\newcommand\atlas{\mathcal{U}}

\newcommand\Ball{\mathbf{B}}

\newcommand\CanLagFr[1][*]{\mathrm{L}\mkern-2mu\mathrm{F}\ifx#1*\else^{(#1)}\fi}
\newcommand\CanLagFrp[1][*]{{\CanLagFr_{\kern-0.2em+}\ifx#1*\else^{(#1)}\fi}}

\newcommand\CanLagFrpt[1][*]{\widetilde{\CanLagFr}{}_{\kern-0.2em+}\ifx#1*\else^{(#1)}\fi}
\newcommand\CanLagFrpyn[1][*]{{\CanLagFr_{\kern-0.2em(+)}\ifx#1*\else^{(#1)}\fi}}
\newcommand\CC{\mathbf{C}}

\newcommand\chit{\widetilde{\chi}}

\newcommand\Ct{\widetilde{C}}
\newcommand\deltak{\delta_k}
\newcommand\deltakt{\widetilde{\delta}_k}
\newcommand\deltaLk{\delta_k^{\mathrm{L}}}
\newcommand\deltaLkt{\deltat_k^{\mathrm{L}}}
\newcommand\deltat{\widetilde{\delta}}

\newcommand\eexp{{\operatorname{e}}}

\newcommand\extder{\mathrm{d}}
\newcommand\Fr{\mathcal{F}}
\newcommand\Frt[1]{\widetilde{\Fr#1}}

\newcommand\ft{\widetilde{f}}

\newcommand\gammat{\widetilde{\gamma}}
\newcommand\Gl{\mathrm{Gl}}
\newcommand\Glk{\Gl_k(n,\CC)}
\newcommand\Glkd{\Gl_k^{(2)}(n,\CC)}
\newcommand\GlnC{\Gl(n,\CC)}
\newcommand\gt{\widetilde{g}}

\newcommand\halfP{$-\tfrac12$-$P$}
\newcommand\HDens[1][]{\Delta\mkern-1mu^{#1}}
\newcommand\HForm[1][]{\widetilde{\Delta}\mkern-1mu^{#1}}
\newcommand\Hilbert{\mathcal{H}}
\newcommand\Hildens[2]{(\mkern-2.5mu( #1,#2)\mkern-2.5mu)}
\newcommand\Hildenst[2]{(\mkern-2.5mu( #1,#2)\mkern-2.5mu)\rlap{$\,\tilde{ }$}}
\newcommand\Hilprod[2]{\langle\mkern-2mu\langle #1,#2\rangle\mkern-2mu\rangle}

\newcommand\ie{i.e.}
\newcommand\LagFr{\mathcal{L}\mkern0mu\mathcal{F}}
\newcommand\LagFrp{\LagFr_+}
\newcommand\LagFrpt{\widetilde{\LagFrp}{}}
\newcommand\LagFrpyn{\LagFr_{(+)}}
\newcommand\LagFrt{\widetilde{\LagFr}}
\newcommand\Li{\mathrm{Liouv}}
\newcommand\Liealg [1]{\mathfrak{#1}}

\newcommand\mapob{\ }
\newcommand\mo{^{-1}}
\newcommand\Ml{\mathrm{Ml}}
\newcommand\Mlk{\Ml_k(n,\CC)}
\newcommand\Mlkd{\Ml_k^{(2)}(n,\CC)}
\newcommand\MlnC{\Ml(n,\CC)}
\newcommand\Mp{\mathrm{Mp}}
\newcommand\MpknR{\Mp_k(2n,\RR)}
\newcommand\MpnR{\Mp(2n,\RR)}

\newcommand\myitemize{$\bullet$ }
\newcommand\myquote[1]{``#1''}
\newcommand\Nt{\widetilde{N}}
\newcommand\nut{\widetilde{\nu}}
\newcommand\oneasmat{\mathbf1}
\newcommand\PairCanLagFrbody{\mathrm{L}\kern-0.25em{}^{^{(2)}}\kern-0.05em\mathrm{F}}
\newcommand\PairCanLagFr[1][*]{\PairCanLagFrbody\ifx#1*\else^{(#1)}\fi}
\newcommand\PairCanLagFrp[1][*]{\PairCanLagFr[#1]_{\kern-0.2em+}}
\newcommand\PairCanLagFrpt[1][*]{\smash{\widetilde{\PairCanLagFrbody}}_{\kern-0.2em+}\ifx#1*\else^{(#1)}\fi}
\newcommand\PairCanLagFrpyn[1][*]{\PairCanLagFr[#1]_{\kern-0.2em(+)}}
\newcommand\PairLagFrD[1][*]{{\mathcal{LF}^{(2)}\ifx#1*\else_{\kern-0.2em#1}\fi}}
\newcommand\PairLagFrDt[1][*]{\widetilde{\mathcal{LF}^{(2)}\ifx#1*\else_{#1}\fi}}
\renewcommand\PairLagFrDt[1][*]{{\smash{\widetilde{\mathcal{LF}}}\ifx#1*\else_{\kern-0.2em#1}\fi^{(2)}}}
\newcommand\PairLagFrpD[1][*]{{\mathcal{LF}^{(2)}\ifx#1*_{\kern-0.2em+}\else_{\kern-0.2em+#1}\fi}}
\newcommand\PairLagFrpDt[1][*]{{\smash{\widetilde{\mathcal{LF}}}\ifx#1*_{\kern-0.2em+}\else_{\kern-0.2em+#1}\fi^{(2)}}}
\newcommand\PairLagFrpynD[1][*]{{\mathcal{LF}^{(2)}\ifx#1*_{\kern-0.2em(+)}\else_{\kern-0.2em(+)#1}\fi}}

\newcommand\PairPolP[1]{{\mathcal{F}P^\Delta_{#1}}{}}
\newcommand\PairPolPt[1]{{\widetilde{\mathcal{F}P}{}^\Delta_{#1}}{}}
\newcommand\Pb{{\overline P}}

\newcommand\pit{\widetilde{\pi}}

\newcommand\projection[1]{\textcolor{red}{#1}}
\newcommand\projectiongreen[1]{\textcolor{blue}{#1}}
\newcommand\pushdown{^{\vrule width0pt depth0pt height0pt}}

\newcommand\pileaf{{\mathrm{pr}}}

\newcommand\rhoMl{\rho_{\Ml}\pushdown}
\newcommand\rhoMp{\rho_{\Mp}\pushdown}

\newcommand\piSymp{\pi_\mathcal{S}}
\newcommand\pitSymp{\pit_\mathcal{S}}
\newcommand\pSymp{p_\mathcal{S}\pushdown}
\newcommand\pSymplos{p_\mathcal{S}}

\newcommand\piLag{\pi_\mathcal{L}\pushdown}
\newcommand\piLaglos{\pi_\mathcal{L}}
\newcommand\pitLag{\pit_\mathcal{L}}
\newcommand\pLag{p_\mathcal{L}}
\newcommand\pLagpd{p_\mathcal{L}\pushdown}

\newcommand\piPairLag{\pi_\mathcal{L}^{(2)}}
\newcommand\pitPairLag{\pit_\mathcal{L}^{(2)}}
\newcommand\pPairLag{p_\mathcal{L}^{(2)}}

\newcommand\piPairPol{\pi_{12}^{\Delta}}
\newcommand\pitPairPol{\pit_{12}^{\Delta}}
\newcommand\pitPairPolsame{\pit_{11}^{\Delta}}
\newcommand\pPairPol{p_{12}^{\Delta}}
\newcommand\pPairPolsame{p_{11}^{\Delta}}

\newcommand\pCanLag{p_L\pushdown}
\newcommand\pPairCanLag{p_L^{(2)}}

\newcommand\PreHilbert[1][*]{\mathit{Pre}\Hilbert\ifx*#1\else_{#1}\fi}

\newcommand\preqinp[2]{\langle#1,#2\rangle}
\newcommand\psit{\widetilde{\smash{\psi}\vrule width0pt height1.3ex}}

\newcommand\RR{\mathbf{R}}

\newcommand\scirc{\,{\raise 0.8pt\hbox{$\scriptstyle\circ$}}\,}
\newcommand\sigmat{\widetilde{\sigma}}
\newcommand\Sp{\mathrm{Sp}}
\newcommand\SpknR{\Sp_k(2n,\RR)}
\newcommand\SpnR{\Sp(2n,\RR)}
\newcommand\st{\widetilde{s}}
\newcommand\stresd[1]{\emph{#1}}
\newcommand\stress[1]{{\bf#1}}
\newcommand\SymFr{\mathcal{S}\mkern-1.5mu\mathcal{F}}
\newcommand\SymFrt{\widetilde{\SymFr}}
\newcommand\SymFrtD{\widetilde{\SymFr_D}{}}

\newcommand\ub{\overline{u}}

\newcommand\ut{\widetilde{u}}

\newcommand\vt{\widetilde{v}}

\newcommand\Xit{\widetilde{\Xi}}

\newcommand\zb{\overline{z}}
\newcommand\ZZ{\mathbf{Z}}

\begin{document}
\setdefaultleftmargin{2.5em}{2.1em}{1.87em}{1.7em}{1em}{1em}

\author{Gijs M. Tuynman}

\title{The metaplectic correction in geometric quantization}

\address[Gijs M. Tuynman]{Laboratoire Paul Painlev\'e, U.M.R. CNRS 8524 et UFR de Math\'ematiques,
Universit\'e de Lille I, 59655 Villeneuve d'Ascq Cedex, France}
\email{FirstName[dot]LastName[at]univ-lille1[dot]fr}

\begin{abstract}
Let $P$ be a polarization on a symplectic manifold for which there exists a metalinear frame bundle.
We show that for any other compatible polarization $P'$ there exists a unique metalinear frame bundle such that the BKS-pairing is well defined.
This means that we do not need the metaplectic frame bundle (nor a positivity condition on $P$) to achieve this  goal, and thus the name ``metaplectic correction'' is inappropriate.

\end{abstract}


\maketitle

\section{Introduction}

In the half-form version of geometric quantization one introduces the bundle of metaframes associated to a polarization $P$ on a symplectic manifold $(M,\omega)$ and then the (complex line) bundle of \halfP-forms as an associated bundle.
In general neither existence nor uniqueness of such a metalinear frame bundle (and its associated line bundle of \halfP-forms) is guaranteed.
Two problems then are faced: (i) how to define a scalar product with these \halfP-forms and (ii) how to relate the Hilbert spaces obtained by two different polarizations (the BKS-pairing).
For problem (ii), in order to be able to integrate over a well defined manifold, one assumes that the two polarizations are compatible (which says that they define a foliation of constant rank whose space of leaves has the structure of a manifold for which the canonical projection is a submersion).
With this condition (and the approach to a solution taken via the BKS kernel), problem (i) turns out to be a particular case of problem (ii).
As said, neither existence nor uniqueness of the metalinear frame bundle is guaranteed, so for two polarizations (or even a whole family of polarizations), our problems mount as to how to guarantee existence of metalinear frame bundles and (once we have existence) how to choose them.
It is generally said or suggested that to do so in a coherent way one needs the metaplectic frame bundle and that one needs to restrict attention to positive (or more precisely, non-negative) polarizations.

What will be shown in this paper is that in order to achieve this goal there is actually no need for the metaplectic frame bundle, nor for the positivity condition.
Once we know a metalinear frame bundle for a single polarization, then for any other compatible polarization there exists a unique metalinear frame bundle for which the BKS-pairing is well defined.
This means that if we have a whole family of polarizations for which we want to compare the quantizations obtained by geometric quantization (in the half-form version), then we only need to specify a single metalinear frame bundle, all others will then be determined uniquely. 
As the freedom in the choice of a metaplectic frame bundle is the same as for a metalinear frame bundle, we won't gain anything by using the metaplectic frame bundle to obtain these metalinear structures.
On the contrary, not using the metaplectic frame bundle allows us to drop the condition that the polarizations should be positive.\footnote{On the other hand, positivity might be needed to guarantee that the Hilbert space constructed by geometric quantization does not reduce to zero, see \cite[p172,174]{Wo91} or \cite[Thm2.8]{Rosenberg1985}}
And it is even conceivable that a metaplectic frame bundle does not exists, whereas for a given polarization there does exist a metalinear frame bundle.

In the first half of this paper we will give an explicit construction for this unique metalinear frame bundle in terms of the transition functions of the initial metalinear frame bundle. 
In the second half of this paper we will show how the metaplectic frame bundle fits into this picture.
We will not provide a full description of the geometric quantization procedure, we will only recall those ingredients needed for our argument. The missing details can be found in all standard texts on geometric quantization (e.g., \cite{Sn80,Wo80,Wo91,Tu85}).

\begin{definition}{A word on notation}
In the sequel we will be confronted with more than a dozen projection maps, all of which one has a tendency to denote by the generic symbol $\pi$. 
As this is highly confusing, especially when several different projections appear in a single formula, we adopt the following conventions.
\begin{enumerate}
\item
Projections (homomorphisms) between (Lie) groups, will be denotes by $\rho$ with a subscript added to distinguish them.

\item
Projections from a fiber bundle to the base space will be denoted by $\pi$ with sub- and superscripts to distinguish the various bundle projections.

\item
When a bundle incorporates ``meta'' objects (metalinear or metaplectic), we will add a twiddle over the $\pi$, and thus use the symbol $\pit$, again with sub- and superscripts.

\item
Projections between various bundles (with meta to without meta) will be denoted by the symbol $p$, again with sub- and superscripts.

\end{enumerate}
On the other hand, when no confusion is possible, we will usually omit the indices in order to improve readability!
At the end of this paper in \S\ref{sectionofallprojections}, the reader will find a summary of all projections used.

\end{definition}

\section{Polarizations, \texorpdfstring{\halfP}{halfP}-densities and \texorpdfstring{\halfP}{halfP}-forms}

In all that will follow, $(M,\omega)$ denotes a connected symplectic manifold of dimension $2n$ with symplectic form $\omega$.
Moreover, we will extend, for each $m\in M$, the skew-symmetric (real) bilinear map $\omega_m:T_mM\times T_mM\to \RR$ by complex bilinearity to a complex bilinear map $\omega_m:T_m^\CC M \times T_m^\CC M\to \CC$.

\begin{definition}{Definitions}
Let $(M,\omega)$ be a symplectic manifold and $D_m\subset T_mM$ a subspace of the tangent space at $m\in M$. We then define the \stresd{symplectic orthogonal $D_m^\perp \subset T_mM$} as
\begin{moneq}
D_m^\perp = \{\,X\in T_m \mid \forall Y\in D_m: \omega(X,Y) = 0\,\}
\mapob.
\end{moneq}
If $D\subset TM$ is a subbundle, we define the subbundle $D^\perp$ as
$
(D^\perp)_m = (D_m)^\perp
$.

A subspace\slash subbundle $D$ is called \stresd{isotropic} if we have the inclusion $D\subset D^\perp$; it is called \stresd{coisotropic} if we have the inclusion $D^\perp\subset D$; and it is called \stresd{Lagrangian} if we have the equality $D^\perp=D$ (which thus is the same as being both isotropic and coisotropic).

\end{definition}

\begin{proclaim}{Lemma}
For any subspace $D_m\subset T_mM$ we have $\dim(D_m^\perp) = 2n-\dim(D_m)$.
And thus in particular any Lagrangian subspace has dimension $n=\dim(M)/2$.

\end{proclaim}

\begin{definition}[defofLagrangianframeandPolarization]{Definitions}
A \stresd{(complex) Langrangian frame} at $m\in M$ is a set $u_1,\dots, u_n$ of $n$ independent elements in the complexified tangent space $T_m^\CC M$ such that
$$
\forall i,j=1, \dots, n\quad:\qquad \omega(u_i,u_j)=0
\mapob.
$$
A \stresd{(complex) Lagrangian subspace} at $m\in M$ is a subspace of $T_m^\CC M$ generated by a Lagrangian frame.
A \stresd{(complex) Lagrangian distribution} $P$ is a (smooth) subbundle of the complexified tangent bundle $T^\CC M$ such that $P_m\subset T_m^\CC M$ is a Lagrangian subspace.
A Lagrangian distribution thus has (constant) rank $n$.
Associated to a Lagrangian distribution $P$ we have its frame bundle $\Fr P$ whose fibres $\Fr P_m$ consist of all bases (over $\CC$) of $P_m$.
This is in a natural way a principal $\GlnC$ bundle when we define the right-action of $\GlnC$ by
\begin{moneq}[defofrightGlnCactiononLagrangiansforP]
(u_1, \dots, u_n)\cdot A \equiv u\cdot A = v \equiv (v_1, \dots, v_n)
\qquad\text{with }
v_j = \sum_{i=1}^n u_i \cdot A_{ij}
\mapob,
\end{moneq}
where $u=(u_1, \dots, u_n)$ is a basis of $\Fr P_m$ and $A\in \GlnC$.

\end{definition}

\begin{definition}[defofcompatibleLagdistributionsandPolarizations]{Definitions}
Let $P_1$ and $P_2$ be two (complex) Lagrangian distributions on $M$. We will say that \stresd{$P_1$ and $P_2$ are compatible} if there exists a (real) distribution $D\subset TM$, whose rank we denote by $k$, such that ${\Pb_1}\cap {P_2} = D^\CC$.

A (complex) Lagrangian distribution $P$ on $M$ is called a \stresd{polarization} if it satisfies the following conditions.
\begin{enumerate}
\item
$P$ is involutive.

\item
$\Pb\cap P$ has constant rank, which implies that there exists a foliation $P_r\subset TM$ such that $\Pb\cap P = P_r^\CC$.

\item
$M/P_r$ admits the structure of a manifold for which the canonical projection is a submersion.

\item
$\Pb + P$ is involutive.

\end{enumerate}
We will say that \stresd{two polarizations $P_1$ and $P_2$ are compatible} if they are compatible as Lagrangian distributions with the additional condition that $M/D$ admits the structure of a manifold for which the canonical projection is a submersion (note that for two polarizations $D$ automatically is involutive).

\end{definition}

\begin{definition}{Nota Bene}
As said, we will denote the rank of $D$ (its dimension) by $k$. A certain number of objects that will follow will depend upon this number.
However, as adding the dependence on this $k$ in the notation will make some of our formul{\ae} more like Christmas trees than mathematics, we will not always make this dependence explicit.

\end{definition}

\begin{definition}{Remarks}
There seems to be no consensus on terminology, as one also finds the names \myquote{(strongly) admissible polarization} for what I here simply call a \myquote{polarization.} 
In those cases the notion of a polarization lacks some of the conditions given here. 
But in the end there is no difference, as one applies geometric quantization only to those $P$ that satisfy all of the conditions given above. So I preferred to skip the intermediate notions and additional adjectives and use the name \myquote{polarization} for those $P$ that satisfy all relevant conditions.

The (main) results presented in this paper are valid for compatible Lagrangian distributions (without any integrability conditions attached). However, it should be noted that the only application is to geometric quantization, where one applies them to compatible polarizations, whose additional (integrability and topological) conditions are needed to prove results that we only allude to.

\end{definition}

\begin{definition}{Definition}
The \stresd{metalinear group} $\MlnC$ is the connected double covering group of $\GlnC $.
It can be realized as the subgroup of $\GlnC \times \CC^*$ by
$$
\MlnC  = \{\, (A,z)\in \GlnC \times \CC^* \mid \det(A) = z^2\,\}
$$
with the obvious projection homomorphism $\projection\rho:\MlnC \to \GlnC $ given by $\projection\rho(A,z) = A$.
If we have to distinguish this group homomorphism from other ones that are also denoted by $\projection\rho$, we will add the subscript $\Ml$ and write $\projectiongreen\rhoMl$.

\end{definition}

\begin{definition}{Definition}
Let $P$ be a Lagrangian distribution and $\Fr P$ the corresponding frame bundle. 
A \stresd{metalinear frame bundle for $P$} is a principal $\MlnC $-bundle $\Frt{P}$ over $M$ together with a bundle map $p:\Frt P\to \Fr P$ such that the following diagram is commutative:
$$
\begin{CD}
\Frt P \times \MlnC  @>>> \Frt P
\\
@V\projection{p\times \rho} VV @VV\projection pV
\\
\Fr P \times \GlnC  @>>> \Fr P\rlap{\mapob,}
\end{CD}
$$
in which the horizontal arrows denote the (right) group actions on these principal bundles.
It follows that the projection\slash bundle map $p:\Frt P\to \Fr P$ is a double covering.
In general, neither existence nor uniqueness of a metalinear frame bundle is guaranteed.
The obstruction to existence is a cohomology class in $H^2(M,\ZZ/2\ZZ)$ determined by the bundle $\Fr P$ and, if we have existence, the inequivalent choices are parametrized by $H^1(M,\ZZ/2\ZZ)$.

\end{definition}

\begin{definition}[densitiesandforms]{A reminder}
Let $X$ be an arbitrary manifold of dimension $d$ and let $\Fr X\to X$ be its (complex) frame bundle, \ie, $\Fr_xX$ consists of all bases (over $\CC$) of $T_x^\CC X$. 
$\Fr X$ is in the obvious way a principal $\Gl(d,\CC)$-bundle and, for $r\in \RR$, an $r$-density on $X$ is a function $W:\Fr X\to \CC$ satisfying the condition
$$
\forall x\in X\ \forall u\in \Fr_x X \ \forall A\in \Gl(d,\CC)
\quad:\qquad
W(u\cdot A) = \vert \det(A)\vert^r \cdot W(u)
\mapob.
$$
The set of all $r$-densities on $X$ can be identified with the set of all sections of a complex line bundle over $X$, associated to the principal $\Gl(d,\CC)$-bundle $\Fr X$ by the representation $\Gl(d,\CC)\to\Gl(\CC)\cong\CC^*$ of $\Gl(d,\CC)$ on $\CC$ given by $A\mapsto \vert\det(A)\vert^{-r}$.

Convergence problems aside, any 1-density $W$ on $X$ can be integrated over $X$ to yield a number $\int_X W$.
The official construction of this number goes as follows.
One chooses an atlas $\{U_\alpha \mid \alpha\in I\}$ and a partition of unity $\{\rho_\alpha\mid\alpha\in I\}$ associated to this atlas. 
And then one defines $\int_X W$ by
$$
\int_X W = \sum_{\alpha\in I}\ \int_{U_\alpha} \ 
\rho_\alpha\bigl(\varphi_\alpha\mo(y)\bigr)\cdot W_{\varphi_\alpha\mo(y)}\bigl( \partial_1\restricted_{\varphi_\alpha\mo(y)}, \dots, \partial_d\restricted_{\varphi_\alpha\mo(y)}\bigr)
\ \extder \lambda^{(d)}(y)
\mapob,
$$
where $\lambda^{(d)}$ denotes the Lebesgue measure in $\RR^d$, where $\varphi_\alpha:U_\alpha\to O_\alpha\subset \RR^d$, $\varphi_\alpha(x) = (y_1, \dots, y_d)$ provides a local coordinate system and where $\partial_i \equiv \partial/\partial y_i$ form everywhere on $U_\alpha$ a basis for the tangent space.
That the result is independent of the choice of the partition of unity and the chosen atlas is a direct consequence of the change of variables formula for the Lebesgue measure and the behavior of a 1-density under the $\Gl(d,\RR)\subset \Gl(d,\CC)$ action: both change with the absolute value of the Jacobian.

If we omit the absolute value in the definition of a 1-density, we get the definition of a \stresd{(volume) form} as being a function $V:\Fr X\to \CC$ satisfying the condition
$$
\forall x\in X\ \forall u\in \Fr_x X \ \forall A\in \Gl(d,\CC)
\quad:\qquad
V(u\cdot A) =  \det(A) \cdot V(u)
\mapob.
$$
The set of all (volume) forms on $X$ can be identified with the set of all sections of a complex line bundle over $X$, associated to the principal $\Gl(d,\CC)$-bundle $\Fr X$ by the representation $\Gl(d,\CC)\to\Gl(\CC)\cong\CC^*$ of $\Gl(d,\CC)$ on $\CC$ given by $A\mapsto \det(A)\mo$.
Moreover, it is not hard to show that this bundle is (isomorphic to) the bundle of (complexified) $d$-forms $\bigwedge^dT^{*\CC}X$ over $X$. 
A (volume) form thus is the same as a (complex) differential form of top degree.

Integration of a volume form $V$ over $X$ is not well defined unless $X$ is orientable, and if it is, its integral depends upon the choice of an orientation.
When those conditions are satisfied, the definition of $\int_X V$ is given by the same formula as for a 1-density, except that one has to use an atlas in which every local coordinate system is oriented positively. In that case all Jacobians will be positive and the absence of the absolute value in the behavior of a (volume) form becomes moot.

\end{definition}

\begin{definition}[defofhalfdensitiesbundle]{Definition}
Let $P$ be a Lagrangian distribution.
A \stresd{\halfP-density} (to be compared with the definition of an $r$-density \recalt{densitiesandforms}) is a function $\nu:\Fr P\to \CC$ satisfying the condition
$$
\forall m\in M\ \ \forall u\in \Fr P_m \ \forall A\in \GlnC \quad:\quad
\nu(u\cdot A) = \vert\det(A)\vert^{-1/2}\cdot \nu(u)
\mapob.
$$
The set of all \halfP-densities can be seen as the set of all sections of a complex line bundle $\HDens[P]M$ over $M$ associated to the principal $\GlnC $-bundle $\Fr P$ by the representation $\GlnC \to \Gl(\CC)\cong\CC^*$ of $\GlnC $ on $\CC$ given by $A\mapsto {\vert\det(A)\vert}^{1/2}$.
The complex line bundle $\HDens[P]M\to M$ is called \stresd{the bundle of \halfP-densities}.

A \stresd{\halfP-form} (to be compared with the definition of a (volume) form  \recalt{densitiesandforms}) is a function $\nut:\Frt P\to \CC$ satisfying the condition
$$
\forall m\in M\ \ \forall\, \ut\in \Frt P_m \ \forall (A,z)\in \MlnC \quad:\quad
\nut\bigl(\ut\cdot (A,z)\bigr) = z\mo\cdot \nut(\ut)
\mapob.
$$
As $z$ is one of the two solutions for $\sqrt{\det(A)}$, this can be suggestively rephrased as
$$
\nut\bigl(\ut\cdot (A,z)\bigr) = \det(A)^{-1/2}\cdot \nut(\ut)
\mapob.
$$
The set of all \halfP-forms can be seen as the set of all sections of a complex line bundle $\HForm[P]M$ over $M$ associated to the principal $\MlnC $-bundle $\Frt P$ by the representation $\MlnC \to \Gl(\CC)\cong\CC^*$ of $\MlnC $ on $\CC$ given by $(A,z)\mapsto z$.
The complex line bundle $\HForm[P]M\to M$ is called \stresd{the bundle of \halfP-forms}.

\end{definition}

\section{Half-density quantization and the function \texorpdfstring{$\deltak$}{delta2}}
\label{halfdensityquantwithmiracles}

An extremely short summary of the half-density version of geometric quantization is the following.
Starting with the symplectic manifold one constructs (if possible) a complex line bundle $L$ (the so called prequantum line bundle) with connection $\nabla$ and a compatible hermitian structure such that the curvature of $\nabla$ equals $-i\omega/\hbar$.
Next one chooses a polarization $P$ and one considers the complex line bundle $L\otimes \HDens[P]M$, the tensor product of $L$ with the complex line bundle of \halfP-densities $\HDens[P] M$.
On $L\otimes \HDens[P]M$ one defines a partial connection $\nabla$ (partial, because it can be defined only for tangent vectors in $P+\overline P$) and one constructs a Hilbert space out of sections of $L\otimes \HDens[P]M$ that are covariantly constant in the direction of $P$.
The construction of the scalar product on this Hilbert space and the attempt to relate the Hilbert spaces corresponding to two different compatible polarizations $P_1$ and $P_2$ follows the same procedure.
It is this procedure that we will now describe in slightly more detail.

We thus assume that $P_1$ and $P_2$ are two compatible polarizations.
We also assume that we have two smooth sections $\psi_i$ of $L\otimes \HDens[P_i]M$ of the form $\psi_i = s_i \otimes \nu_i$, where $s_i$ is a (smooth) section of the prequantum line bundle $L$ and where $\nu_i$ is a (smooth) section of $\HDens[P_i]M$.
The first step then is to construct a map that to each $m\in M$ associates a 1-density $\Hildens{\psi_1}{\psi_2}_m$ at $\projectiongreen\pileaf(m) \in M/D$, where $\projectiongreen\pileaf:M\to M/D$ denotes the canonical projection of $M$ onto the leaf space $M/D$.
The next step is to show that, if the $\psi_i$ are covariantly constant in the direction of $P_i$, then this 1-density is independent of the choice of $m$ as long as $\projectiongreen\pileaf(m)$ is unchanged.
This implies that we have created a 1-density $\Hildens{\psi_1}{\psi_2}$ on $M/D$, which might be integrable over this space.
And then the argument splits into two, depending upon whether we have $P_1=P_2$ or not.

When we have $P_1=P_2=P$, the 1-density $\Hildens{\psi_1}{\psi_2}$ on $M/D$ is used to define the Hilbert space and its scalar product. 
One starts with the vector space $\PreHilbert[P]$ of sections $\psi$ of $L\otimes \HDens[P]M$ that are convariantly constant in the direction of $P$ and for which $\int_{M/D} \Hildens{\psi}{\psi}<\infty$:
\begin{align*}
\PreHilbert[P] 
&
= \Bigl\{\,\psi:M\to L\otimes \HDens[P]M \text{ smooth} \Bigm\vert
\\
&\kern6em
\forall X\in P: \nabla_X\psi = 0 \ \&\ \int_{M/D} \Hildens{\psi}{\psi}<\infty
\,\Bigr\}
\mapob.
\end{align*}
On this vector space one defines the scalar product $\Hilprod{\ }{\ }$ by
$$
\Hilprod{\psi_1}{\psi_2} = \int_{M/D} \Hildens{\psi_1}{\psi_2}
\mapob,
$$
and then one defines the Hilbert space $\Hilbert_P$ as the completion of the pre-Hilbert space $\bigl(\PreHilbert[P], \Hilprod{\ }{\ }\,\bigr)$.

When we have $P_1\neq P_2$ (and in that context one speaks about the BKS-pairing, after R.J.~Blattner, B.~Kostant and S.~Sternberg who introduced this pairing), one first assumes that the Hilbert spaces $\Hilbert_{P_i}$ are already defined. And then one hopes for the existence of a unique isomorphism (a unitary complex linear bijection) $\Phi:\Hilbert_{P_1} \to \Hilbert_{P_2}$ and a constant $C\in\CC^*$ such that we have the equality
\begin{moneq}[arbitraryconstantinBKS]
\Hilprod{\Phi(\psi_1)}{\psi_2}_{\Hilbert_{P_2}} = C\cdot \int_{M/D} \Hildens{\psi_1}{\psi_2}
\end{moneq}
for all $\psi_i \in \Hilbert_{P_i}$ for which the right hand side is well defined, \ie, for which the $1$-density $\Hildens{\psi_1}{\psi_2}$ is integrable over $M/D$.
Unfortunately (to the best of my knowledge), there does not exist a useful criterion that tells us when this will happen.
In some simple cases this is indeed the case, in particular for $M=\RR^{2n}$ with the vertical, horizontal and holomorphic polarizations, for which this BKS-pairing reproduces the Fourier transform or the Bargmann transform.
On the other hand, there exist examples for which the BKS-pairing does define a linear map $\Phi$, but one which is not unitary.

\medskip

So far the general theory. We will now concentrate on the details of the first step, \ie, the construction of the 1-density $\Hildens{\psi_1}{\psi_2}_m$ at $\projectiongreen\pileaf(m) \in M/D$, as it is this step that will provide the clue to our claims.
We start with some definitions that will be used throughout.

\begin{definition}[defofGlkandGlkd]{Definition}
For $0\le k\le n$ we define the subgroup $\Glk\subset \GlnC$ as the subgroup of those elements $g\in \GlnC$ of the form
$$
g=\begin{pmatrix} A & B \\ \mathbf0 & D \end{pmatrix}
\mapob,
$$
where $A\in \Gl(k,\RR)$, $D\in \Gl(n-k,\CC)$ and $B$ an arbitrary complex matrix of the appropriate size.
$\Glk$ thus is the subgroup that preserves the \stress{real} subspace generated by the first $k$ elements of the canonical basis of $\CC^n$.
We also introduce the group $\Glkd\subset \Glk^2$ of elements $(g_1,g_2)\in \Glk^2$ of the form
\begin{moneq}[specialformGkld]
(g_1,g_2)
=
(\, \begin{pmatrix} {A} & B_1 \\ \mathbf0 & D_1 \end{pmatrix} 
\,,\,
\begin{pmatrix} {A} & B_2 \\ \mathbf0 & D_2 \end{pmatrix} \, )
\mapob.
\end{moneq}

\end{definition}

\begin{definition}[defofPairPolanddeltak]{Definitions}
Let $P_i$, $i=1,2$ be two compatible Lagrangian distributions and $\Fr P_i$ the corresponding frame bundles. We then define the bundle $\PairPolP{12}$ by defining its fibres as
$$
\PairPolP{12}\caprestricted_m
=
\{\, (u,v)\in \Fr P_1\caprestricted_m \times \Fr P_2\caprestricted_m \mid
\forall 1\le i\le k : \ub_i = u_i = v_i\,\}
\mapob.
$$
$\PairPolP{12}$ thus is the subbundle of the product bundle $\Fr P_1\times_M \Fr P_2$ consisting of those couples $(u,v)$ in which $u$ is a basis for $P_1\caprestricted_m$, $v$ a basis for $P_2\caprestricted_m$ and $u_1$, \dots, $u_k$ a basis for $D_m$ (not $D_m^\CC$, that is why we added the condition $u_i=\ub_i$).

On $\PairPolP{12}$ we define the function $\deltak:\PairPolP{12}\to \CC$ by
\begin{moneq}[definitionofdelta2]
(u,v)\in \PairPolP{12}\caprestricted_m
\qquad\Longrightarrow\qquad
\deltak(u,v) = \det\bigl(\,-i\cdot \omega(\ub_i, v_j)_{i,j=k+1}^n\,\bigr)
\mapob.
\end{moneq}
The factor $-i$ in front of $\omega$ is for the moment purely artificial, especially when one knows that we will take, in this section, the absolute value of $\deltak$. 
However, later on this factor will avoid some awkward (but unimportant) factors. 
Changing this factor (or others like it) will only change the constant $C$ used in \recalf{arbitraryconstantinBKS}, no other result will be affected by such a change.

\end{definition}

\begin{proclaim}[behaviourdeltakunderGlkd]{Lemma}
The bundle $\PairPolP{12}$ is a principal fibre bundle over $M$ with structure group $\Glkd$, $\deltak$ takes values in $\CC^*$ and for all $(u,v)\in \PairPolP{12}\caprestricted_m$ and all $(g_1,g_2)\in \Glkd$ of the form \recalf{specialformGkld} we have
\begin{align}
\deltak\bigl( (u,v)\cdot (g_1,g_2)\bigr)
&
=
\deltak(u,v) \cdot \overline{\det(D_1)}\cdot{\det(D_2)}
\notag
\\
&
=
\deltak(u,v) \cdot \overline{\det(g_1)}\cdot{\det(g_2)} \cdot \det(A)^{-2}
\mapob.
\label{tranformationpropdeltak}
\end{align}

\end{proclaim}

\begin{definition}{Definition}
Let $P_i$, $i=1,2$ be two compatible Lagrangian distributions, let $\psi_i$, $i=1,2$ be a section of $L\otimes\HDens[P_i]M$ of the form $\psi_i = s_i\otimes \nu_i$ with $s_i$ a section of $L$ and $\nu_i$ a section of $\HDens[P_i] M$ and let $m\in M$ be arbitrary. Then we define the 1-density $\Hildens{\psi_1}{\psi_2}_m$ at $\projectiongreen\pileaf(m) \in M/D$ by the formula
\begin{align}
\Hildens{\psi_1}{\psi_2}_m
(w)
&
=
\preqinp{ s_1(m)}{s_2(m)}\cdot  \overline{\nu_1(u)}
\cdot  
{\nu_2(v)\vrule width0pt height2.0ex} 
\notag
\\
&\kern3em
\cdot 
\sqrt{\bigl\vert\deltak(u,v)\bigr\vert\vrule width0pt height 1.78ex}
\cdot 
\vert\Li(u_1, \dots, u_k, W)\vert
\mapob,
\label{defofHildensforhalfdensities}
\end{align}
where $w\in \Fr (M/D)$ is an arbitrary basis of $T_{\projectiongreen\pileaf(m)}^\CC (M/D)$, 
where $(u,v)\in \PairPolP{12}\caprestricted_m$ is arbitrary, where $W_1, \dots, W_{2n-k}\in T_m^\CC M$ are such that they project to the frame $w$ at $\projectiongreen\pileaf(m)$:
\begin{moneq}
\forall 1\le i\le 2n-k
:
\projectiongreen\pileaf_*(W_i) = w_i
\mapob,
\end{moneq}
and where $\Li$ is the Liouville volume form on $M$ defined by
$$
\Li = \frac{(-1)^{n(n-1)/2}}{n!}\cdot \omega^n
\mapob.
$$

\end{definition}

\begin{proclaim}[Hildensfordensitiesindepofuv]{Lemma}
The 1-density $\Hildens{\psi_1}{\psi_2}_m$ is (indeed) independent of the choice of $(u,v)\in \PairPolP{12}\caprestricted_m$.

\end{proclaim}

When we look at the definition \recalf{defofHildensforhalfdensities} of the 1-density $\Hildens{\psi_1}{\psi_2}_m$, we see first of all terms that are quite natural: the hermitian form on $L$ applied to the two sections $s_1$ and $s_2$ and the two \halfP-densities $\nu_1$ and $\nu_2$. 
At the end we find another relatively natural term: the Liouville volume form.
This is a volume form defined on the (symplectic) manifold $M$, not on the quotient space $M/D$, but the completion of the basis $w$ of $T^\CC_{\projectiongreen\pileaf(m)}$ (or better, its lift to vectors $W_i$) to a basis of $T^\CC_mM$ with the vectors $u_1, \dots, u_k$ which span $D$ seems natural, especially given that we used these same vectors in the frames on which the $\nu_i$ are evaluated.
Moreover, the dependence on the vectors $w$ (or their lifts to $M$) is such that $\vert\Li\vert$ indeed behaves as a 1-density (due to the use of the absolute value) when changing by an element in $\Gl(2n-k,\CC)$.
Moreover, as the freedom in the vectors $W_i$ is the addition of an element of $D$, the fact that we added the basis vectors $u_1, \dots, u_k$ for $D$ in the Liouville form implies that the result is independent of the choice of these vectors $W_i$.
Remains the term $\vert\deltak(u,v)\vert^{1/2}$.

Given the other terms, this factor can easily be explained in two steps.
Obviously the other terms depend upon the choice of the frames $(u,v)\in \PairPolP{12}\caprestricted_m$. Now if we change to another element in $\PairPolP{12}\caprestricted_m$, this is done by an element $(g_1,g_2)$ of $\Glkd$.
And then $\nu_1(u)$ changes with a factor $\vert\det(g_1)\vert^{-1/2}$, $\nu_2(v)$ changes with a factor $\vert\det(g_2)\vert^{-1/2}$ and the Liouville 1-density (a 1-density because of the absolute value) changes with the factor $\vert\det(A)\vert$.
So if we want the total to be independent of such a choice, the missing factor should depend upon the couple $(u,v)$ in such a way that it changes with the inverse factor, \ie, with the factor 
$$
\vert\det(g_1)\vert^{1/2}\cdot\vert\det(g_2)\vert^{1/2} \cdot \vert\det(A)\vert\mo
=
\bigl\vert\overline{\det(g_1)}\cdot {\det(g_2)} \cdot \det(A)^{-2} \bigr\vert^{1/2}
\mapob.
$$
And according to \recalt{behaviourdeltakunderGlkd}, the function $\vert\deltak(u,v)\vert^{1/2}$ does just that (note that this is a proof of \recalt{Hildensfordensitiesindepofuv}).

\section{Half-form quantization and the function \texorpdfstring{$\deltakt$}{deltakt}}
\label{halfformquantization}

The half-form version of geometric quantization follows from start to finish the same scheme as the half-density version, except that the bundle of \halfP-densities is replaced by the bundle of \halfP-forms, provided such a bundle exists.
More precisely, one replaces the complex line bundle $L\otimes\HDens[P]M$ by the bundle $L\otimes\HForm[P]M$, the tensor product of the prequantume line bundle $L$ with the bundle $\HForm[P] M$ of \halfP-forms.
On this bundle one defines a partial connection $\nabla$ (partial, but now defined only for elements of $P$) and the Hilbert space then is constructed out of sections of $L\otimes\HForm[P]M$ that are covariantly constant in the direction of $P$, just as in the half-density version.
And, just as in the half-density version, the construction of the scalar product and the BKS-pairing starts with the definition of a 1-density $\Hildenst{\psit_1}{\psit_2}_m$ at $\projectiongreen\pileaf(m) \in M/D$ associated to two sections $\psit_i$ of $L\otimes\HForm[P_i]M$ of the form $\psit_i = s_i \otimes \nut_i$ with $\nut_i$ a section of $\HForm[P_i]M$ (for two compatible polarizations $P_1$ and $P_2$).
The next step is to show that, if the $\psit_i$ are covarianty constant in the direction of $P_i$, then this 1-density is independent of the choice of $m$ as long as $\projectiongreen\pileaf(m)$ is unchanged.
This implies that we have created a 1-density $\Hildenst{\psit_1}{\psit_2}$ on $M/D$, which might be integrable over this space.
Starting at this point, the argument for the half-density version is copied word by word.

And thus again the crucial point is the construction of the 1-density $\Hildenst{\psit_1}{\psit_2}_m$ at $\projectiongreen\pileaf(m) \in M/D$.
As the absolute value of the determinant no longer intervenes in the definition of \halfP-forms, one is tempted to give the following definition, adapting the formula for the half-density case by leaving out (some of) the absolute values:
\begin{align}
\Hildenst{\psit_1}{\psit_2}_m
(w)
&
=
\preqinp{ s_1(m)}{s_2(m)}\cdot  \overline{\nut_1(\ut)\vrule width0pt height2.0ex} \cdot {\nut_2(\vt)\vrule width0pt height2.0ex} 
\cdot
\sqrt{\deltak(u,v)}
\notag
\\&
\kern3em
\cdot \vert\Li(u_1, \dots, u_k, W)\vert
\mapob,
\label{tentativeHildensforms}
\end{align}
where $w\in \Fr(X/D)$ is an arbitrary basis of $T_{\projectiongreen\pileaf(m)}^\CC (M/D)$, 
where $\ut\in \Frt{P_1}\caprestricted_m$ and $\vt\in \Frt{P_2}\caprestricted_m$ are arbitrary with the restriction that $\bigl(\projectiongreen\pileaf_*(\ut), \projectiongreen\pileaf_*(\vt)\bigr) = (u,v) \in \PairPolP{12}\caprestricted_m$, and where $W_1, \dots, W_{2n-k}\in T_m^\CC M$ are such that they project to the frame $w$ at $\projectiongreen\pileaf(m)$.

This idea works quite well, except that we have some sign problems: first of all we do not know which square root to take of $\deltak(u,v)$. And changing the choice of the metaframes $\ut$ or $\vt$ will introduce a sign (via the functions $\nut_i$).
The idea then is to replace the (undefined) square root of $\deltak(u,v)$ by a function $\deltakt(\ut,\vt)$ depending upon the metaframes in such a way that we won't have any sign problems.
To obtain necessary and sufficient conditions for this to be possible, we need some definitions, analogous to the definitions \recalt{defofGlkandGlkd} and \recalt{defofPairPolanddeltak}.

\begin{definition}{Definition}
For $0\le k\le n$ we define the subgroup $\Mlk\subset \MlnC$ as the inverse image of $\Glk$ under the homomorphism $\MlnC\to \GlnC$.
We also introduce the group $\Mlkd\subset \Mlk^2$ as the inverse image of $\Glkd$ under the homomorphism $\projection\rho\times\projection\rho$. Its elements $(\gt_1,\gt_2)$ thus are of the form
\begin{moneq}[structuregroupPairPoltNEW]
(\gt_1,\gt_2)
=
\Bigl(\quad
(\, \begin{pmatrix} {A} & B_1 \\ \mathbf0 & D_1 \end{pmatrix} , z_1\, )
\quad,\quad
(\, \begin{pmatrix} {A} & B_2 \\ \mathbf0 & D_2 \end{pmatrix} , z_2\, )
\quad\Bigr)
\mapob,
\end{moneq}
with $A\in \Gl(k,\RR)$, $D_i\in \Gl(n-k,\CC)$, $B_i$ arbitrary complex matrices of the appropriate size and $z_i^2 = \det(A)\cdot \det(D_i)$.

\end{definition}

\begin{definition}{Definition}
Let $P_i$, $i=1,2$ be two compatible Lagrangian distributions and $\Fr P_i$ the corresponding frame bundles. 
Assume that we have metalinear frame bundles $\projection p_i:\Frt{P_i} \to\Fr{ P_i}$.
We then define the bundle $\PairPolPt{12}$ as the subbundle of $\Frt{P_1}\times_M\Frt{P_2}$ that projects to $\PairPolP{12}$:
\begin{align*}
\PairPolPt{12}\caprestricted_m 
&
=
\bigl\{\,(\ut,\vt)\in \Frt{P_1}\caprestricted_m \times \Frt{P_2}\caprestricted_m \mid
\bigl(\projection p_1(\ut), \projection p_2(\vt)\bigr) \in \PairPolP{12}\caprestricted_m\,\bigr\}
\\
\noalign{\vskip2\jot}
&
=
\Bigl\{\,(\ut,\vt)\in \Frt{P_1}\caprestricted_m \times \Frt{P_2}\caprestricted_m \mid
\forall 1\le i\le k : \overline{\projection p_1(\ut)\vrule width0pt height1.95ex}_i = {\projection p_1(\ut)}_i = {\projection p_2(\vt)\vrule width0pt height1.95ex}_i
\,\Bigr\}
\mapob.
\end{align*}
It follows immediately that the projection $\projectiongreen\pPairPol:\PairPolPt{12}\to \PairPolP{12}$ defined as
\begin{moneq}
\projectiongreen\pPairPol(\ut,\vt) = \bigl(p_1(\ut), p_2(\vt)\bigr)
\end{moneq}
is a $4$-$1$ covering map.

\end{definition}

\begin{proclaim}{Lemma}
The bundle $\PairPolPt{12}$ is a principal fibre bundle over $M$ with structure group $\Mlkd$.

\end{proclaim}

With these preparations we now can copy our heuristic arguments used at the end of section \ref{halfdensityquantwithmiracles}.
Given $\psit_i=s_i\otimes \nut_i$ it seems natural to have $\preqinp{ s_1(m)}{s_2(m)}$ as a factor in the definition of $\Hildenst{\psit_1}{\psit_2}_m$.
The product $\overline{\nut_1(\ut)\vrule width0pt height2.0ex} \cdot {\nut_2(\vt)}$ also seems natural. 
Using the vectors $u_1, \dots, u_k$ spanning $D$ in the Liouville volume form is as natural here as it was in the half-density version.
And using the absolute value of the Liouville volume form gives the right property of a 1-density at $\projectiongreen\pileaf(m)$ when changing the basis $w$ at $\projectiongreen\pileaf(m)$.

So let us see how these terms change when we change the couple $(\ut,\vt)\in \PairPolPt{12}\caprestricted_m$ with an element $(\gt_1,\gt_2)\in \Mlkd$.
First of all the term with $\nut_1$ changes with a factor $(\zb_1)\mo$ and the term with $\nut_2$ changes with a factor $z_2\mo$. 
And the term with the Liouville volume form changes with the factor $\vert\det(A)\vert$.
The \myquote{missing} term thus should change with the factor
\begin{moneq}[behaviourmissingfactorhalfforms]
\zb_1\cdot {z_2}\cdot \vert\det(A)\vert\mo
\mapob.
\end{moneq}
As this looks quite like the behaviour of the square root of $\deltak$, without the absolute value (remember, $z_i^2 = \det(g_i)$ and $\det(A)$ is real), it thus becomes natural to look for a smooth function $\deltakt:\PairPolPt{12}$ satisfying the conditions 
\begin{align}
\bigl(\deltakt(\ut,\vt)\bigr)^2 
&
= \deltak\bigl(\projection p_1(\ut),\projection p_2(\vt)\bigr)
\label{deltaktrootofdeltak}
\\
\noalign{\vskip2\jot}
\deltakt\bigl((\ut,\vt)\cdot(\gt_1,\gt_2)\bigr) 
&
= \deltakt(\ut,\vt) \cdot \overline{z_1}\cdot {z_2}\cdot \vert\det(A)\vert\mo
\label{behaviourdeltaktunderMlkd}
\end{align}
for all $(\ut,\vt)\in \PairPolPt{12}$ and all $(\gt_1,\gt_2)\in \Mlkd$.

\begin{definition}{Remark}
It is tempting to think that \recalf{deltaktrootofdeltak} implies \recalf{behaviourdeltaktunderMlkd}, given the behaviour of the function $\deltak$ and the fact that we require $\deltakt$ to be smooth. This is indeed true on the connected component containing the identity in $\Mlkd$, on which $\det(A)>0$. 
But $\Mlkd$ has two connected components because $\Gl(k,\RR)$ does. So the behaviour on the other component is not determined by \recalf{deltaktrootofdeltak}, but might contain a minus sign.

On the other hand, the distribution $D$ is orientable if and only if the bundle $\PairPolPt{12}$ has two connected components. And if that is the case, we can restrict attention to one of these components (the choice of an orientation for $D$) and reduce the structure group to those elements in $\Mlkd$ with $\det(A)>0$.
And then indeed the property \recalf{deltaktrootofdeltak} (and smoothness of $\deltakt$) implies \recalf{behaviourdeltaktunderMlkd} for this (reduced) structure group.
But when $D$ is not orientable, there is no way to choose the transition functions in the reduced structure group (\ie, with $\det(A)>0$), and then we need the additional condition \recalf{behaviourdeltaktunderMlkd} to determine which square root we have to take on the other component in a given fiber. 

\end{definition}

\begin{definition}{Definition}
Let $P_1$ and $P_2$ be two compatible Lagrangian distributions and let $\Frt{P_1}$ and $\Frt{P_2}$ be metalinear frame bundles for $P_1$ and $P_2$ respectively. We will say that the metalinear frame bundles $\Frt{P_1}$ and $\Frt{P_2}$ are \stresd{compatible} if there exists a smooth function $\deltakt:\PairPolPt{12}\to \CC^*$ satisfying the conditions \recalf{deltaktrootofdeltak} and \recalf{behaviourdeltaktunderMlkd}.

\end{definition}

\begin{proclaim}[locallyonlytwochoicesfordeltakt]{Lemma}
Let $U\subset M$ be connected. If $\deltakt, \deltakt':\projectiongreen\pitPairPol\mo(U) \subset \PairPolPt{12}\to\CC^*$ (with $\projectiongreen\pitPairPol:\PairPolPt{12}\to M$ the projection onto the base space) are two smooth functions satisfying \recalf{deltaktrootofdeltak} and \recalf{behaviourdeltaktunderMlkd}, then necessarily $\deltakt'=\pm \deltakt$.

\end{proclaim}

\begin{proclaim}[localexistenceofholwortNEW]{Lemma}
If $U\subset M$ is contractible, then there exists a smooth function $\deltakt:\projectiongreen\pitPairPol\mo(U)\subset \PairPolPt{12}\to \CC^*$ satisfying the conditions \recalf{deltaktrootofdeltak} and \recalf{behaviourdeltaktunderMlkd}.

\end{proclaim}

\begin{preuve}
As $U$ is contractible, there exists a trivializing section $\st:U\to \PairPolPt{12}$.
Hence the composite smooth function $\gamma:U\to \CC^*$ given by 
$$
\gamma(m) = \deltak\bigl(\, \projectiongreen\pPairPol\bigl(\st(m)\bigr)\,\bigr)
$$
is defined on a contractible set, so there exists a smooth function $\gammat:U\to \CC^*$ such that $\gammat^2 = \gamma$.
Moreover, the local section $\st$ defines a diffeomorphism $\Phi:U\times \Mlkd\to \projectiongreen\pitPairPol\mo(U)$ given by
$$
\Phi\bigl(m,(\gt_1,\gt_2)\bigr) = \st(m)\cdot (\gt_1,\gt_2)
\mapob.
$$
We now define the function $\deltakt:\projectiongreen\pitPairPol\mo(U)\to \CC^*$ by
$$
(\deltakt\scirc \Phi)\bigl(m,(\gt_1,\gt_2)\bigr) = \gammat(m) \cdot {\overline{z_1} \cdot z_2}\cdot{\vert\det(A)\vert}\mo
\mapob,
$$
using the expression \recalf{structuregroupPairPoltNEW} for $(\gt_1,\gt_2)$.
It then is a straightforward computation to show that this $\deltakt$ satisfies the conditions \recalf{deltaktrootofdeltak} and \recalf{behaviourdeltaktunderMlkd}. 
\end{preuve}

\begin{definition}{Definition}
We will say that an open cover $(U_\alpha)_{\alpha\in I}$ is a \stresd{nice cover} if any finite intersection of elements of the cover is either empty or contractible.
For an arbitrary open cover there always exists a nice cover that is a refinement (each element of the nice cover is included in an element of the original cover). It suffices to choose a metric and then to consider geodesically convex open subsets.
The advantage of using nice covers is that any locally trivial fibre bundle is automatically trivial on any element of a nice cover and that any closed form is exact on an element of a nice cover.

\end{definition}

\begin{proclaim}[uniquenessofsecondmetalinearbundleNEW]{Theorem}
Let $P_1$ and $P_2$ be compatible Lagrangian distributions.
Then for any metalinear frame bundle $\Frt{P_1}$ for $P_1$ there exists a unique metalinear frame bundle $\Frt{P_2}$ for $P_2$ that is compatible with $\Frt{P_1}$.

\end{proclaim}

\begin{preuve}
We start by choosing a nice cover $(U_\alpha)_{\alpha\in I}$ of trivializing charts for $\PairPolP{12}$ with trivializing sections $s_\alpha:U_\alpha\to \PairPolP{12}$.
The associated transition functions $g\AB$ are (necessarily) of the form
$$
g\AB
=
\bigl(\, g\AB^{(1)}, g\AB^{(2)} \,\bigr)
=
\Bigl(\quad
\begin{pmatrix} {A}\AB & B^{(1)}\AB \\ \mathbf0 & D^{(1)}\AB \end{pmatrix} 
\quad,\quad
\begin{pmatrix} {A\AB} & B^{(2)}\AB \\ \mathbf0 & D^{(2)}\AB \end{pmatrix} 
\quad\Bigr)
\mapob.
$$
The local sections $s_\alpha$ immediately determine, by projection on the components of the product $\PairPolP{12}\subset \Fr{P_1} \times_M \Fr{P_2}$, trivializing sections $s_\alpha^{(i)}$ for the bundles $\Fr{P_i}$.
Moreover, the transition functions associated to these sections are exactly the functions $g\AB^{(i)}$.

To prove uniqueness, we assume that we have a metalinear frame bundle $\Frt{P_1}$ for $P_1$ and two metalinear frame bundles $\Frt{P_2}$ and $\Frt{P_2}{}'$ for $P_2$. And we assume that we have globally defined smooth functions $\deltakt:\PairPolPt{12} \to \CC^*$ and $\deltakt':\PairPolPt{12}'  \to \CC^*$ satisfying the conditions \recalf{deltaktrootofdeltak} and \recalf{behaviourdeltaktunderMlkd}.
And then we have to show that the metalinear frame bundles $\Frt{P_2}$ and $\Frt{P_2}{}'$ are equivalent.

As the $U_\alpha$ are contractible, there exist sections $\st_\alpha^{(1)}$, and $\st_\alpha^{(2)}$, $\st_\alpha^{(2)}{}'$ of $\Frt{P_1}$, $\Frt{P_2}$, and $\Frt{P_2}{}'$ respectively that project onto the sections $s_\alpha^{(1)}$, $s_\alpha^{(2)}$, and $s_\alpha^{(2)}$ (sic!).
And then the functions $\st_\alpha=(\st_\alpha^{(1)}, \st_\alpha^{(2)})$ and $\st{}'_\alpha=(\st_\alpha^{(1)}, \st_\alpha^{(2)}{}')$ are trivializing sections of $\PairPolPt{12}$ and $\PairPolPt{12}'$.
According to the construction, the transition functions $\gt\AB^{(\prime)}\in \Mlkd$ associated to these trivializations are necessarily of the form
$$
\gt\AB^{(\prime)}
=
\bigl(\,\gt\AB^{(1)}, \gt\AB^{(2)}{}^{(\prime)} \,\bigr)
=
\Bigl(\ 
(\, \begin{pmatrix} {A}\AB & B^{(1)}\AB \\ \mathbf0 & D^{(1)}\AB \end{pmatrix} , z^{(1)}\AB\, )
\ ,\ 
(\, \begin{pmatrix} \,\overline{A}\AB & B^{(2)}\AB \\ \,\mathbf0 & D^{(2)}\AB \end{pmatrix} , z^{(2)}\AB{}^{(\prime)}\, )
\ \Bigr)
\mapob,
$$
for suitable functions $z\AB^{(1)}, z\AB^{(2)}, z\AB^{(2)}{}':U_\alpha\cap U_\beta\to \CC^*$.

We now define the functions $\deltat_\alpha, \deltat_\alpha':U_\alpha\to \CC^*$ by
$$
\deltat_\alpha(m) = \deltakt\bigl(\st_\alpha(m)\bigr)
\qquad\text{and}\qquad
\deltat_\alpha'(m) = \deltakt'\bigl(\st_\alpha'(m)\bigr)
\mapob.
$$
By construction of the sections $\st_\alpha^{(\prime)}$ and the fact that the functions $\deltakt^{(\prime)}$ satisfy \recalf{deltaktrootofdeltak}, we have the equalities
$$
\bigl( \deltat_\alpha(m) \bigr)^2 = \deltak\bigl(s_\alpha(m)\bigr) = \bigl( \deltat_\alpha'(m) \bigr)^2
\mapob.
$$
By connectedness of $U_\alpha$ it then follows that there exist constants $\epsilon_\alpha =\pm1$ such that 
$$
\forall m\in U_\alpha
\quad:\quad
\deltat_\alpha'(m)  = \epsilon_\alpha\cdot \deltat_\alpha(m)
\mapob.
$$
We now note that, if we change the section $\st_\alpha^{\,\prime}$ by the element $(\oneasmat,\epsilon_\alpha)\in \Mlk\subset\MlnC$, then the function $\deltat_\alpha'$ changes by a factor $\epsilon_\alpha$ according to \recalf{behaviourdeltaktunderMlkd}.
It follows that we may suppose that we have $\deltat_\alpha'=\deltat_\alpha$.

On the other hand, by \recalf{behaviourdeltaktunderMlkd} and the form of the transition functions, we have the equalities
\begin{align*}
\deltat_\beta(m) 
&
= \deltakt\bigl(\st_\alpha(m)\cdot \gt\AB(m)\bigr)
=
\deltat_\alpha(m) \cdot \overline{z\AB^{(1)}} \cdot {z\AB^{(2)}} \cdot \bigl\vert\det\bigl( A\AB(m)\bigr)\bigr\vert\mo
\\
\noalign{\vskip2\jot}
\deltat_\beta'(m) 
&
= \deltakt'\bigl(\st_\alpha^{\,\prime}(m)\cdot \gt\AB'(m)\bigr)
=
\deltat_\alpha'(m) \cdot \overline{z\AB^{(1)}} \cdot {z\AB^{(2)}{}'} \cdot \bigl\vert\det\bigl( A\AB(m)\bigr)\bigr\vert\mo
\mapob,
\end{align*}
and thus, taking the quotient of these two equalities, we obtain
$
{z\AB^{(2)}{}'} =  {z\AB^{(2)}}
$,
\ie, $\Frt{P_2}$ and $\Frt{P_2}{}'$ are equivalent.

\bigskip

The proof of existence follows the same ideas, but now we assume only the existence of a metalinear frame bundle $\Frt{P_1}$ for $P_1$, and we have to find a metalinear frame bundle $\Frt{P_2}$ for $P_2$ and a smooth function $\deltakt:\PairPolPt{12}\to \CC^*$ satisfying \recalf{deltaktrootofdeltak} and \recalf{behaviourdeltaktunderMlkd}.
To that end we start by looking at the functions $\delta_\alpha:U_\alpha\to \CC^*$ defined as
$$
\delta_\alpha(m) = \deltak\bigl(s_\alpha(m)\bigr)
\mapob.
$$
Using these, we define the functions $g_\alpha:U_\alpha\to \Glkd$ by
$$
g_\alpha(m) = 
(\, \begin{pmatrix} \oneasmat & \mathbf0 \\ \mathbf0 & D_\alpha(m) \end{pmatrix}
\ ,\ 
\begin{pmatrix} \oneasmat & \mathbf0 \\ \mathbf0 & \oneasmat \end{pmatrix} \, )
$$
with $D_\alpha(m)$ a diagonal matrix with one diagonal element equal to $\delta_\alpha(m)$ and all others equal $1$.
It then follows that the (trivializing) sections $s_\alpha':U_\alpha\to \PairPolP{12}$ defined as
$$
s_\alpha'(m) = s_\alpha(m) \cdot \bigl(g_\alpha(m)\bigr)\mo
$$
have the property that (using \recalf{behaviourdeltakunderGlkd})
$$
\deltak\bigl(s_\alpha'(m)\bigr) 
=
\delta_\alpha(m) \cdot \det\bigl(D_\alpha(m)\bigr)\mo 
=
\delta_\alpha(m) \cdot \bigl(\delta_\alpha(m)\bigr)\mo
=
1
\mapob.
$$
The upshot of this computation is that we may assume without loss of generality that the functions $\delta_\alpha$ are identically~$1$.
But with this assumption, we can make the computation, again using \recalf{behaviourdeltakunderGlkd}
\begin{align}
1
&
=\deltak\bigl( s_\beta(m)\bigr)
=
\deltak\bigl(s_\alpha(m)\cdot g\AB(m) \bigr)
\notag
\\
&
=
\deltak\bigl(s_\alpha(m)\bigr) \cdot \overline{\det\bigl( g\AB^{(1)} \bigr)} \cdot {\det\bigl( g\AB^{(2)} \bigr)} \cdot \det(A\AB)^{-2}
\notag
\\
&
=
\overline{\det\bigl( g\AB^{(1)} \bigr)} \cdot {\det\bigl( g\AB^{(2)} \bigr)} \cdot \det(A\AB)^{-2}
\notag
\\
&
=
\overline{\det\bigl( g\AB^{(1)} \bigr)} \cdot {\det\bigl( g\AB^{(2)} \bigr)} \cdot \vert\det(A\AB)\vert^{-2}
\mapob.
\label{squareofmetalinearframebundletransitionfunctionNEW}
\end{align}

Now by hypothesis there exists a metalinear frame bundle $\Frt{P_1}$, hence, as the trivializing charts $U_\alpha$ are contractible, there also exist local trivializing sections $\st_\alpha^{(1)}:U_\alpha\to \Frt{P_1}$ that project to the trivializing sections $s_\alpha^{(1)}$ of $\Fr{P_1}$.
By definition of a metalinear frame bundle, the associated transition functions $\gt^{(1)}\AB:U_\alpha\cap U_\beta\to \MlnC$ are given by
$$
\gt\AB^{(1)}
=
(\, g\AB^{(1)} , z^{(1)}\AB\, )
\mapob,
$$
where the $z\AB^{(1)}$ are smooth functions satisfying 
$$
\bigl(z\AB^{(1)}(m)\bigr)^2
=
\det\bigl(g\AB^{(1)}(m)\bigr)
\mapob.
$$
It follows, using \recalf{squareofmetalinearframebundletransitionfunctionNEW}, that the functions $z\AB^{(2)}$ defined by
\begin{moneq}[defoftransitionfuncformetalinearcompatibleNEW]
z\AB^{(2)}(m)
=
\bigl\vert\det\bigl(A\AB(m)\bigr)\bigr\vert \cdot \overline{{z\AB^{(1)}(m)}\mo}
\mapob,
\end{moneq}
satisfy the condition $\bigl(z\AB^{(2)}(m)\bigr)^2
=
\det\bigl(g\AB^{(2)}(m)\bigr)$.
Moreover, as the $A\AB$ and $z\AB^{(1)}$ satisfy the cocycle condition, so do these $z\AB^{(2)}$.
And hence the functions $\gt^{(2)}\AB:U_\alpha\cap U_\beta\to \MlnC$ defined by
$$
\gt\AB^{(2)}
=
(\, g\AB^{(2)} , z^{(2)}\AB\, )
$$
are the transition functions of a metalinear frame bundle $\Frt{P_2}$ for $P_2$.

We now recall that the construction of the (principal) fibre bundle $\Frt{P_2}$ via the transition functions and the (trivializing) cover $(U_\alpha)_{\alpha\in I}$ automatically gives us local trivializing sections $\st_\alpha^{(2)}:U_\alpha\to \Frt{P_2}$.
Moreover, it is immediate from the construction that the sections $\st_\alpha:U_\alpha\to \Frt{P_1}\times \Frt{P_2}$ defined as 
$$
\st_\alpha(m) = \bigl(\st_\alpha^{(1)}(m), \st_\alpha^{(2)}(m)\bigr)
$$
are actually (local, trivializing) sections of the subbundle $\PairPolPt{12}$.

With these preparations we can define a smooth function $\deltakt$ on $\projectiongreen\pitPairPol\mo(U_\alpha)$ by
$$
\deltakt\bigl( \st_\alpha(m)\cdot (\gt_1,\gt_2) \bigr)
=
\overline{z_1}\cdot {z_2} \cdot \vert\det(A)\vert\mo
\mapob, 
$$
where we used the expression \recalf{structuregroupPairPoltNEW} for an element of the structure group of the principal bundle $\PairPolPt{12}$.
That these local definitions coincide on overlaps is a direct consequence of the defining property \recalf{defoftransitionfuncformetalinearcompatibleNEW} of the transition functions.
To finish, we note that by construction this $\deltakt$ satisfies \recalf{behaviourdeltaktunderMlkd}. That it also satisfies \recalf{deltaktrootofdeltak} is a direct consequence of the hypothesis that we have $\deltak\scirc s_\alpha = 1$.
\end{preuve}

\begin{definition}{Definition}
Let $P_i$, $i=1,2$ be two compatible Lagrangian distributions, 
let $\Frt{P_i}$ be two compatible metalinear frame bundles for $P_1$ and $P_2$ respectively,
let $\psit_i$, $i=1,2$ be a section of $L\otimes\HForm[P_i]M$ of the form $\psit_i = s_i\otimes \nut_i$ with $s_i$ a section of $L$ and $\nut_i$ a section of $\HForm[P_i] M$ and let $m\in M$ be arbitrary. Then we define the 1-density $\Hildenst{\psit_1}{\psit_2}_m$ at $\projectiongreen\pileaf(m) \in M/D$ by the formula
\begin{align}
\Hildenst{\psit_1}{\psit_2}_m
(w)
&
=
\preqinp{ s_1(m)}{s_2(m)}\cdot  \overline{\nut_1(\ut)\vrule width0pt height2.0ex}
\cdot  {\nut_2(\vt)\vrule width0pt height2.0ex} \cdot 
\deltakt(\ut,\vt)
\notag
\\&
\kern3em
\cdot 
\vert\Li(u_1, \dots, u_k, W_{1}, \dots, W_{2n-k})\vert
\mapob,
\label{densityonMDforhalfforms}
\end{align}
where $w$ is an arbitrary basis of $T_{\projectiongreen\pileaf(m)}^\CC (M/D)$, 
where $(\ut,\vt)\in \PairPolPt{12}\caprestricted_m$ is arbitrary with $u=\projectiongreen\pileaf(\ut)$, 
and where $W_{1}, \dots, W_{2n-k}\in T_m^\CC M$ are such that they project to the frame $w$ at $\projectiongreen\pileaf(m)$.

\end{definition}

\begin{proclaim}[Hildensforformsindepofchoiceutvt]{Lemma}
The 1-density $\Hildenst{\psit_1}{\psit_2}_m$ is (indeed) independent of the choice of $(\ut,\vt)\in \PairPolPt{12}\caprestricted_m$.

\end{proclaim}

Once we have this $1$-density on $M/D$, we can follow any text on geometric quantization to show that if the $\psit_i$ are covariantly constant in the directions of $P_i$, then $\Hildenst{\psit_1}{\psit_2}_m$ is independent of $m$ for $\projectiongreen\pileaf(m)$ fixed. We thus have a well-defined $1$-density on $M/D$.
This $1$-density is the basis for the BKS-pairing as well as for the scalar products on the respective Hilbert spaces. 
But\dots, in order to be able to define the scalar product we need one more result: a metalinear frame bundle for a polarization $P$ should be compatible with itself, simply because we want to use two (covariantly constant) sections $\psit_1$ and $\psit_2$ of \textbf{the same} metalinear frame bundle!

\begin{proclaim}[metalinearframebundlecompatiblewithitself]{Proposition}
Let $P$ be a polarization and $\Frt{P}$ a metalinear frame bundle for $P$. Then $\Frt{P}$ is compatible with itself.

\end{proclaim}

\begin{preuve}
To prove that $\Frt{P_1}$ is compatible with itself, we have to exhibit a smooth function $\deltakt:\PairPolPt{11}\subset \Frt{P_1} \times_M \Frt{P_1}\to \CC^*$ satisfying the conditions \recalf{deltaktrootofdeltak} and \recalf{behaviourdeltaktunderMlkd}.
So let $(U_\alpha)_{\alpha\in I}$ be a trivializing nice cover for $\PairPolPt{11} $. 
If $\st_\alpha:U_\alpha\to \PairPolPt{11} $ is a local section, then we may assume without loss of generality that it is of the form
$$
\st_\alpha(m) = \bigl(\sigmat_\alpha(m), \sigmat_\alpha(m)\bigr)
$$ 
for some trivializing section $\sigmat_\alpha:U_\alpha\to \Frt{P_1}$.
It then follows that the transition functions $\gt\AB:U_\alpha\cap U_\beta \to \Mlkd$ are of the form
$$
\gt\AB(m) = \bigl(\chit\AB(m), \chit\AB(m)\bigr)
$$
with
$$
\chit\AB(m) = (\chi\AB(m), z\AB(m))
=
\bigl(\  \begin{pmatrix} A\AB(m) & B\AB(m) \\ \mathbf0 & D\AB(m)  \end{pmatrix}   \ ,\  z\AB(m)\ \bigr)\in \Mlk
$$
and that the $\chit\AB$ are the transition functions for $\Frt{P_1}$.
To facilitate the coming computations, we introduce the functions $\delta_\alpha:U_\alpha\to \CC^*$ by
$$
\delta_\alpha(m) = 
\deltak\bigl( \, \projectiongreen \pPairPolsame\bigl(\st_\alpha(m)\bigr) \,\bigr)
\mapob,
$$
for which we have the property (for $m\in U_\alpha\cap U_\beta)$
\begin{align}
\delta_\beta(m)
&
=
\delta_\alpha(m) \cdot \overline{\det\bigl(\chi\AB(m)\bigr)} \cdot {\det\bigl(\chi\AB(m)\bigr)}\cdot \det\bigl(A\AB(m)\bigr)^{-2}
\notag
\\
&
=
\delta_\alpha(m) \cdot 
\bigl(\  \overline{z\AB(m)}\cdot{z\AB(m)}\cdot \vert\det\bigl(A\AB(m)\bigr)\vert\mo\ \bigr)^2
\mapob.
\label{transformationofdeltaalpha}
\end{align}
We next note that (for any $\alpha$ and any $m\in U_\alpha$) the element $\projectiongreen \pPairPolsame\bigl(\st_\alpha(m)\bigr)$ is of the form $\projectiongreen \pPairPolsame\bigl(\st_\alpha(m)\bigr)=(u,u)$ for some $u\in \Fr P$.
It follows that we have
$$
\delta_\alpha(m) = \det\bigl( \,-i\cdot\omega(\ub_i, u_j)_{i,j=k+1}^n\bigr)
\mapob.
$$
Now the matrix $-i\cdot\omega(\ub_i, u_j)_{i,j=k+1}^n$ is hermitian, so its determinant is real (and it is non-zero). 
Hence the $\delta_\alpha$ are of constant sign. It then follows immediately from \recalf{transformationofdeltaalpha} that this sign does not depend upon $\alpha$.
Hence there exists $\varepsilon\in \{0,1\}$ such that, for all $\alpha$ and all $m\in U_\alpha$, we have
\begin{moneq}[constantsignofralpha]
\delta_\alpha(m) = \eexp^{\varepsilon\pi i}\cdot \vert  \delta_\alpha(m) \vert
\mapob.
\end{moneq}

With these preparations we define the function $\deltakt$ on $\projectiongreen\pitPairPolsame\mo(U_\alpha)$ by
$$
\deltakt\bigl( \st_\alpha(m)\cdot \gt\,\bigr)
=
\eexp^{\varepsilon\pi i/2}\cdot \vert  \delta_\alpha(m) \vert^{1/2}
\cdot
\zb_1\cdot z_2 \cdot \vert\det(A)\vert\mo
\mapob,
$$
where $\gt\in \Mlkd$ is of the form \recalf{structuregroupPairPoltNEW}.
That these local definitions coincide on overlaps is a direct consequence of \recalf{transformationofdeltaalpha} and \recalf{constantsignofralpha}.
That this $\deltakt$ satisfies \recalf{deltaktrootofdeltak} and \recalf{behaviourdeltaktunderMlkd} is immediate from its definition (and the definition of the functions $\delta_\alpha$).
\end{preuve}

\begin{definition}[changingdeltaktbyaconstant]{Remark}
When we want to use the $1$-density $\Hildenst{\psit_1}{\psit_2}_m$ associated to two sections of the same bundle $L\otimes\HForm[P]M$ to define a scalar product on these sections, we need (at least) that the $1$-density $\Hildenst{\psit}{\psit}_m$ is positive (when we use the same section at both slots).
This is not necessarily the case for the $1$-density using the function $\deltat_k$ as defined in the proof of \recalt{metalinearframebundlecompatiblewithitself}.
However, looking at that proof, it is immediate that when we multiply it by $\eexp^{-\varepsilon\pi i/2}$, then the result will be positive when using the same section $\psi$ in both slots.
As we already know that $\deltakt$ is unique only up to a global sign \recalt{locallyonlytwochoicesfordeltakt}, adding another global factor $- i$ should not worry us too much.
The more so when we remember that we allow for an arbitrary (global) factor when comparing two different quantizations (see \recalf{arbitraryconstantinBKS}), a global factor that can be interpreted as changing the function $\deltat_k$ with this global factor.

\end{definition}

\begin{definition}{A word on the orbit method}
Let $G$ be a connected Lie group, $\Liealg g$ its Lie algebra and let $\mathcal{O}_{\mu_o}$ be the coadjoint orbit of $G$ through $\mu_o\in \Liealg g^*$.
A $G$-invariant polarization on $\mathcal{O}_{\mu_o}$ is described by a subalgebra $\Liealg h\subset \Liealg g^\CC$ satisfying some conditions.
When one applies the half-form version of geometric quantization to this situation, the interesting object is not the metalinear frame bundle itself, but a metalinear frame bundle to which the action of the group can be lifted.
Those metalinear frame bundles are parametrized by characters $\chit:G_{\mu_o}\to \CC^*$ satisfying $\chit(g)^2 = \det\bigl(\operatorname{Ad}_{\Liealg h/\Liealg g_{\mu_o}^\CC}(g)\bigr)$, where $G_{\mu_o}$ is the stabilizer subgroup of $\mu_o$.
Our result then takes the following form: if $\Liealg h_1$ and $\Liealg h_2$ are two compatible polarizations and if there exists such a character for $\Liealg h_1$, then there exists a unique character for $\Liealg h_2$ for which the BKS-pairing is well defined.
If $\chit_1:G_{\mu_o}\to \CC^*$ is the character for the first polarization, then the character for the second is given by $\chit_2(g) = \det_{\Liealg d/\Liealg g_o}(g)\,/\,\,\overline{\chit_1(g)}$, where the subalgebra $\Liealg d\subset \Liealg g$ is defined by $\Liealg h_1\cap \overline{\Liealg h_2} = \Liealg d^\CC$ and where $\Liealg g_o$ is the Lie algebra of $G_{\mu_o}$.

\end{definition}

\section{A nice idea}

We have seen that if we have a polarization, then (in the half-form version of geometric quantization) we need a metalinear frame bundle in order to define a Hilbert space and a representation by self-adjoint operators of quantizable observables. 
When we want to compare two such representations associated to two different polarizations, we only know a systematic way to do so when these two polarizations are compatible (and even then we need some miracles to happen). And then the knowledge of a metalinear frame bundle for one polarization completely determines the metalinear frame bundles for all other compatible polarizations.
But this means that we first decide which polarization interests us most, and then, starting from a metalinear frame bundle for a fixed polarization, we construct metalinear frame bundles for all other polarizations that are compatible, \ie, for which we know a systematic way to compare the obtained representations.

But wouldn't it be nice if we had a systematic way to obtain a metalinear frame bundle for all polarizations beforehand in such a way that we are guaranteed that, whenever two of the polarizations are compatible, then the corresponding metalinear frame bundles are automatically compatible? 
The following idea explains how we might realize this (details follow later).

We start by defining the bundle $\LagFr M$ of all Lagrangian frames on $M$. If $P$ is a polarization, its frame bundle $\Fr{P}$ is in a natural way a subbundle of $\LagFr M$. Or said differently, $\LagFr M$ is the union (over all Lagrangian distributions $P$) of all frame bundles $\Fr{P}$. The Lagrangian frame bundle $\LagFr M$ has a natural right action of $\GlnC$ (which naturally is compatible with the right action of $\GlnC$ on any frame bundle $\Fr{P}$ associated to a polarization).
Now suppose there exists a double covering $\LagFrt M\to \LagFr M$ with a right-action of $\MlnC$ that is compatible with the $\GlnC$ action on $\LagFr M$.
Then for any polarization $P$ we can take the preimage in $\LagFrt M$ of the frame bundle $\Fr{P}$, seen as subbundle of $\LagFr M$; this preimage then is a metalinear frame bundle $\Frt{P}$ for $P$.
In this way we would have a unified way to obtain metalinear frame bundles for all polarizations simultaneously.

And with the bundle $\LagFrt M$ we can define a subbundle $\PairLagFrDt M\subset \LagFrt M \times_M \LagFrt M$, just as we defined the subbundle $\PairPolPt{12} M\subset \Frt{P_1} \times \Frt{P_2}$.
And, just as $\Frt{P}$ is a subbundle of $\LagFrt M$, the bundle $\PairPolPt{12} M$ is a subbundle of $\PairLagFrDt M$ (for the metalinear frame bundles obtained via $\LagFrt M$).
It thus is tempting to hypothesize that there exists a (global, smooth) function $\deltakt$ on $\PairLagFrDt M$ (abuse of notation justified by what follows) such that its restriction to any subbundle $\PairPolPt{12} M$ is a globally defined function $\deltakt$ satisfying the conditions \recalf{deltaktrootofdeltak} and \recalf{behaviourdeltaktunderMlkd}.
It would follow that the metalinear frame bundles obtained via $\LagFrt M$ are automatically compatible.

Unfortunately, this idea breaks down already at the first stage, as (as far as I know) the bundle $\LagFrt M$ does not exist.
But, as we will explain in the next sections (which is for a (very) large part, including some of the notation, a copy of \cite[p87--97]{Sn80}, even when it is not mentioned explicitly), the metaplectic frame bundle carries out this idea when one restricts attention to \stress{positive} polarizations.
On the other hand, let me stress again that this construction with the metaplectic bundle does not add any relevant information.
For suppose we have a family of positive polarizations, some of which are compatible (and some not).
If there exists a metaplectic bundle, we thus obtain metalinear frame bundles for all these polarizations. And as bonus we know that if two of them are compatible, the obtained metalinear frame bundles will be compatible too.
But given a polarzation $P$ in this family, we could have chosen any metalinear frame bundle for it (it will exist, as we already have one). And by changing the metaplectic frame bundle we could have obtained all possible choices for this metalinear frame bundle. 
So for this single polarization we do not gain anything by using (a choice for) the metaplectic bundle.
Now if $P'$ is another polarization in this family, then there are two possibilities: either it is compatible with $P$ or it is not. If it is compatible, there exists a unique metalinear frame bundle for it that will be compatible with the one chosen for $P$, which will be the one obtained via the metaplectic bundle.
And if it is not compatible, there is no reason to use the same metaplectic bundle to define its metalinear frame bundle, we could have chosen any other metaplectic bundle as well. 
So once again we can obtain all possible choices for its metalinear frame bundle.

\section{The metaplectic frame bundle and typical fibers}

As the details of the constructions become rather technical, we start with a short outline of what will follow in this section.
As said above, the purpose is to define the bundle $\LagFrt M$, or rather $\LagFrpt M$ of Lagrangian metaframes associated to \stress{positive} polarizations (to be defined).
The way one realizes this is by first considering $\LagFr M$ (and its subbundle $\LagFrp M$) as an associated bundle to the principal fiber bundle of symplectic frames $\SymFr M$.
Then to introduce the notion of the metaplectic frame bundle $\SymFrt M\to \SymFr M$ (a principal fiber bundle with structure group the metaplectic group $\MpnR$) and to define $\LagFrpt M$ as an associated bundle to $\SymFrt M$.

In order to carry out this program, we need to define the typical fiber $\CanLagFr$ of $\LagFr M$ and the associated typical fiber $\CanLagFrp\subset \CanLagFr$ of $\LagFrp M$, as well as the typical fiber $\CanLagFrpt$ of $\LagFrpt M$. 
And because these are typical fibers of associated bundles, we need a left-action of the symplectic group $\SpnR$ on $\CanLagFrpyn$ and a left-action of the metaplectic group $\MpnR$ on $\CanLagFrpt$. 
And finally, because we want the bundles $\LagFrp M$ and $\LagFrpt M$ to have a right-action of $\GlnC$ and $\MlnC$ respectiveley, we need a right-action of $\GlnC$ and $\MlnC$ on $\CanLagFrp$ and $\CanLagFrpt$ respectively, right-actions that commute with the left-actions of $\SpnR$ and $\MpnR$ respectively.

\begin{definition}{A technical detail}
The typical fiber $\CanLagFr$ is a (regular) submanifold of the vector space $\GlnC^2\cong \CC^{2n^2}$, but $\CanLagFrp\subset \CanLagFr$ is a submanifold with boundary and corners.
As such, the fiber bundles $\LagFrp M$ and a fortiori $\LagFrpt M$ are not manifolds in the usual sense.
In particular the notion of smoothness of a function on these bundles (and bundles derived from them) is not well defined.
What is guaranteed is that these bundles are in the realm of topological manifolds, and all our maps will be continuous (in particular the map that will generalize $\deltakt$).

On the other hand, all our bundles are defined (can be defined) in terms of trivializing charts and transition functions. And the transition functions will always be smooth functions defined on open subsets of the base manifold $M$. 
In this way we will stay quite close to the notion of ordinary manifolds. 
And more importantly, our generalization of the function $\deltakt$ will be a continuous function whose square will be smooth on subbundles that are regular manifolds.
And hence its restriction to the subbundle in question will be smooth as needed for our notion of compatibility.

Having made this remark, we will make no more mention of this \myquote{detail} in the sequel, as it will not affect our argumentation.

\end{definition}

\begin{definition}{Definitions}
The \stresd{Lagrangian frame bundle} $\LagFr M$ is the bundle over $M$ whose fibres $\LagFr_mM$ consist of all Lagrangian frames at $m\in M$, see \recalt{defofLagrangianframeandPolarization}. 
The bundle $\LagFr M$ has a natural right-action of $\GlnC$: for $u=(u_1, \dots, u_n)\in \LagFr_mM$ and $A\in \GlnC$ we have
\begin{moneq}[defofrightGlnCactiononLagrangians]
u\cdot A = v = (v_1, \dots, v_n)
\qquad\text{with }
v_j = \sum_{i=1}^n u_i \cdot A_{ij}
\mapob.
\end{moneq}
If $P$ is a Lagrangian distribution, its frame bundle $\Fr P$ (whose fibres $\Fr P_m$ consist of all bases (over $\CC$) of $P_m$) is in a natural way a subbundle of $\LagFr M$ in such a way that the natural right-actions of $\GlnC$ on $\Fr{P}$ (see \recalf{defofrightGlnCactiononLagrangiansforP}) and $\LagFr M$ coincide. 

\medskip

A Lagrangian frame $u$ at $m\in M$ is said to be \stresd{positive} (non-negative would be a better but more awkward name) if the hermitian matrix $H\in M(n,\CC)$ defined by
\begin{moneq}[hermitianmatrixforlagrangiansubspace]
H_{ij} = -i\cdot \omega(\ub_i,u_j)
\end{moneq}
has no strictly negative eigenvalues. This condition is equivalent to the condition 
$$
\forall x\in \mathrm{Span}(u_1, \dots, u_n)
\quad:\quad
-i\cdot\omega(\overline{x},{x})\ge0
\mapob.
$$
Associated to the notion of a positive Lagrangian frame we define the subset $\LagFrp M \subset \LagFr M$ as the subset of all \stresd{positive} Lagrangian frames; it has in a natural way the structure of a (sub)bundle over $M$ which is invariant under the right-action of $\GlnC$ (but, as we will see, its typical fiber is a manifold with boundary and corners). 
And a Lagrangian distribution $P$ will be called \stresd{positive} if (for all $m\in M$) the Lagrangian subspace $P_m$ admits a positive frame\slash basis, in which case all frames\slash bases of $P_m$ will be positive.
Note that, if $P$ is a positive Lagrangian distribution, its frame bundle $\Fr P$ is in a natural way a subbundle of $\LagFrp M$.

\medskip

The \stresd{symplectic frame bundle} $\SymFr M$ is the subbundle of all frames (of $TM$) formed by those bases of the tangent space that are \myquote{canonical} with respect to the symplectic form. More precisely, for $m\in M$ the fibre $\SymFr_mM$ consists of those bases
$$
(e;f)\equiv (e_1, \dots, e_n, f_1, \dots, f_n) \in T_mM
$$
satisfying the conditions
$$
\forall i,j=1,\dots,n\quad:\qquad
\omega(e_i,e_j) = 0 = \omega(f_i,f_j)
\qquad\text{and}\qquad
\omega(e_i,f_j) = \delta_{ij}
\mapob.
$$
It is a principal $\Sp(2n,\RR)$-bundle over $M$, where $\Sp(2n,\RR)$ denotes the symplectic group. 
It will sometimes be useful to identify an element $A\in \Sp(2n,\RR)$ with a set of four matrices $T_i\in \Gl(n,\RR)$ as
\begin{moneq}[SpasfourGl]
A = \begin{pmatrix} T_1 & T_2 \\ T_3 & T_4 \end{pmatrix}
\mapob.
\end{moneq}
The condition of belonging to $\Sp(2n,\RR)$ then can be written as the conditions
\begin{moneq}[conditionsonSp]
T_4^t\cdot T_1 - T_2^t\cdot T_3 = \oneasmat_{n}
\kern1.5em,\kern1.5em
T_1^t\cdot T_3 = T_3^t\cdot T_1
\kern1.5em,\kern1.5em
T_2^t\cdot T_4 = T_4^t\cdot T_2
\mapob.
\end{moneq}
For $A\in \Sp(2n,\RR)$ the action on a (symplectic) frame is defined by
$$
(e;f)\cdot A
\equiv
(e;f)\cdot \begin{pmatrix} T_1 & T_2 \\ T_3 & T_4 \end{pmatrix}
=
(\hat e;\hat f)
\quad\text{with}\quad
\left\{
\begin{aligned}
\hat e_j 
&
= \sum_{i=1}^n \bigl(e_i\, (T_1)_{ij} + f_i\,(T_3)_{ij} \bigr)
\\
\hat f_j
&
= \sum_{i=1}^n \bigl(e_i \,(T_2)_{ij} + f_i\,(T_4)_{ij} \bigr)
\mapob.
\end{aligned}
\right.
$$

\end{definition}

\begin{definition}{Definition}
Let $G\to \Sp(2n,\RR)$ be the (connected) universal covering group of $\Sp(2n,\RR)$; its kernel is (isomorphic to) $\ZZ$.
The \stresd{metaplectic group} $\Mp(2n,\RR)$ is defined as the quotient $\Mp(2n,\RR)=G/2\ZZ$.
Elementary algebra then tells us that we have an induced homomorphism $\projectiongreen\rhoMp:\Mp(2n,\RR) \to\Sp(2n,\RR)$ with kernel $\ZZ/2\ZZ$ such that the following diagram is commutative:
$$
\begin{matrix}
G 
\\
\smash{\vrule depth30pt} & \searrow
\\
\noalign{\vskip1\jot}
&& G/2\ZZ \equiv Mp(2n,\RR)
\\
\noalign{\vskip1\jot}
\downarrow & \swarrow\!\rlap{\raise-5pt\hbox{$ \projectiongreen\rhoMp$}}
\\
\noalign{\vskip2\jot}
\llap{$G/\ZZ \equiv$} \,\Sp\rlap{$(2n,\RR)$\mapob.}
\end{matrix}
$$
Nota bene: neither $G$ nor $\Mp(2n,\RR)$ can be realized as a matrix group, which is the same as saying that they don't have finite dimensional faithful representations.

\end{definition}

\begin{definition}[defofmetaplecticbundle]{Definition}
A \stresd{metaplectic frame bundle} over $M$ is a principal $\Mp(2n,\RR)$-bundle $\SymFrt M$ over $M$ together with a bundle map $\projectiongreen\pSymp:\SymFrt M\to \SymFr M$ such that the following diagram is commutative:
$$
\begin{CD}
\SymFrt M \times \Mp(2n,\RR) @>>> \SymFrt M
\\
@V\projectiongreen{\pSymp\times \rhoMp} VV @VV\projectiongreen \pSymp V
\\
\SymFr M \times \Sp(2n,\RR) @>>> \SymFr M\rlap{\mapob,}
\end{CD}
$$
in which the horizontal arrows denote the (right) group actions on these principal bundles.
It follows that projection\slash bundle map $\projectiongreen \pSymp:\SymFrt M\to \SymFr M$ is a double covering.
In general, neither existence nor uniqueness of a metaplectic frame bundle is guaranteed.
As for metalinear frame bundles, the obstruction to existence is a cohomology class in $H^2(M,\ZZ/2\ZZ)$ (this time determined by the bundle $\SymFr M$) and, if we have existence, the inequivalent choices are parametrized by $H^1(M,\ZZ/2\ZZ)$.

\end{definition}

\begin{definition}{Definitions}
The (complex) vector space $M(n,\CC)^2$ admits a natural left-action of $\Sp(2n,\RR)$ (actually of $\Gl(2n,\CC)\supset \Sp(2n,\RR)$ but that is of no importance here) and a natural right-action of $\GlnC $ commuting with the $\Sp(2n,\RR)$-action. For $A=\bigl( \begin{smallmatrix} T_1 & T_2 \\ T_3 & T_4 \end{smallmatrix}\bigr)\in \Sp(2n,\RR)$ (see \recalf{SpasfourGl}) and $C\in \GlnC $ these actions are defined by
$$
A
\cdot 
\begin{pmatrix} U \\ V \end{pmatrix} =
\begin{pmatrix} T_1\cdot U + T_2\cdot V \\ T_3\cdot U + T_4 \cdot V \end{pmatrix}
\qquad\text{and}\qquad
\begin{pmatrix} U \\ V \end{pmatrix} \cdot C
=
\begin{pmatrix} U\cdot C \\ V\cdot C \end{pmatrix}
\mapob.
$$

Let us denote by $a_1, \dots, a_n, b_1, \dots, b_n$ the canonical basis of $\CC^{2n}$ and by $\omega_o$ the canonical symplectic form on $\CC^{2n}$ defined by
$$
\forall i,j=1,\dots,n\quad:\qquad
\omega_o(a_i,a_j) = 0 = \omega_o(b_i,b_j)
\qquad\text{and}\qquad
\omega_o(a_i,b_j) = \delta_{ij}
\mapob.
$$
We interpret the couple $(\CC^{2n}, \omega_o)$ as a model for a complexified tangent space $T_m^\CC M$ with the symplectic form $\omega_m$.
Any set of $n$ vectors $u_1, \dots, u_n$ in $\CC^{2n}$ is determined uniquely by two matrices $U,V\in M(n,\CC)$ according to
$$
u_j = \sum_{i=1}^n (a_i\, U_{ij} + b_i\, V_{ij})
\quad\text{or equivalently}\quad
u=(a;b)\cdot \begin{pmatrix} U\\ V \end{pmatrix}
\mapob.
$$
The vectors $u_1, \dots, u_n$ form a Lagrangian frame in $\CC^{2n}$ if and only if the matrices $U$ and $V$ satisfy the two conditions
\begin{moneq}[conditionsofCanLagFr]
\det(\,U^\dagger \, U + V^\dagger\, V)\neq 0
\qquad\text{and}\qquad
U^t\, V = V^t \, U
\mapob,
\end{moneq}
where the superscript $t$ denotes the transpose and where the superscript $\dagger$ denotes the hermitian conjugate, \ie, complex conjugation and transpose.
The first condition assures that the vectors $u_1, \dots, u_n$ are independent and the second condition assures that the subspace generated by the $u_i$ is isotropic.
Moreover, the Lagrangian frame $u$ is positive if and only if the matrix
\begin{moneq}
i\,(\,V^\dagger\, U - U^\dagger\, V)
\end{moneq}
is non-negative definite, \ie, $x^\dagger\cdot i\,(\,V^\dagger\, U - U^\dagger\, V)\cdot x\ge0$ for all $x\in \CC^n$.

We thus can define the submanifold $\CanLagFr\subset M(n,\CC)^2$ by
$$
\CanLagFr = \{\, (U,V)\in M(n,\CC)^2 \mid \det(U^\dagger\, U + V^\dagger\, V)\neq 0
\ \&\ 
U^t\cdot V = V^t \cdot U
\,\}
\mapob,
$$
which thus is a manifold isomorphic to the set of all Lagrangian frames in $\CC^{2n}$.
By abuse of terminology, we will say that $\CanLagFr$ \myquote{is} the set of all Lagrangian frames in $\CC^{2n}$.
This submanifold of $M(n,\CC)^2$ is (obviously) invariant under the (right) $\GlnC $-action, but it is also invariant under the (left) $\Sp(2n,\RR)$ action: the first condition in \recalf{conditionsofCanLagFr} is preserved because $A\in \Sp(2n,\RR)$ is invertible and the second condition is preserved because of \recalf{conditionsonSp}.

We also define the subset $\CanLagFrp\subset \CanLagFr$ of (or better, corresponding to) non-negative Lagrangian frames by
\begin{moneq}
\CanLagFrp
=
\{\, (U,V)\in \CanLagFr \mid i\,(\,V^\dagger\, U - U^\dagger\, V) \text{ is non-negative definite}\,\}
\mapob.
\end{moneq}
$\CanLagFrp$ (just as $\CanLagFr$) is invariant under the left-action of $\Sp(2n,\RR)$ and the right-action of $\GlnC$.

\end{definition}

\begin{proclaim}[symframegivesbijection]{Lemma}
A symplectic frame $(e;f)\in \SymFr_mM$ determines a bijection $\CanLagFr \to \LagFr_mM $ by
$$
\begin{pmatrix} U\\ V\end{pmatrix} \mapsto u
\qquad\text{with}\qquad
u_j = \sum_{i=1}^n (e_i \,U_{ij} + f_i\,V_{ij})
\quad\text{or}\quad
u=(e;f)\cdot \begin{pmatrix} U\\ V\end{pmatrix}
\mapob.
$$

\end{proclaim}

\begin{proclaim}[LagFrInducedbySymFr]{Lemma}
The bundle $\projectiongreen\piLag:\LagFr M\to M$ has $\CanLagFr$ as typical fiber and is associated to the principal $\Sp(2n,\RR)$-bundle $\projectiongreen\piSymp:\SymFr M\to M$ by the left-action of $\Sp(2n,\RR)$ on $\CanLagFr$.
Moreover, the right-action of $\GlnC$ on $\LagFr M$ corresponds to the right-action of $\GlnC$ on $\CanLagFr$.

\end{proclaim}

\begin{preuve}
The bundle $\pi_B:B\to M$ associated to $\projectiongreen\piSymp:\SymFr M\to M$ by the left-action of $\Sp(2n,\RR)$ on $\CanLagFr$ is defined as the quotient 
\begin{moneq}
B=\SymFr M \times \CanLagFr / \Sp(2n,\RR) \equiv \SymFr M \times_{\Sp(2n,\RR)} \CanLagFr
\end{moneq}
with respect to the $\Sp(2n,\RR)$-action
\begin{moneq}
g\cdot \bigl( (e;f),(U,V)\bigr) = \bigl( (e;f)\cdot g,g\mo\cdot(U,V)\bigr)
\mapob.
\end{moneq}
It then is elementary to show that the map $\Psi:\SymFr M \times \CanLagFr \to \LagFr M$ defined by
\begin{moneq}
\Psi\bigl( (e;f),(U,V)\bigr) = (e;f)\cdot \begin{pmatrix} U\\ V \end{pmatrix}
\end{moneq}
induces a bundle isomorphism $B\to \LagFr M$ which is compatible with the right-actions of $\GlnC$.

The argument using local trivializations is a bit longer, but provides additional information about specific trivializations that will be used later.
So let $(U_\alpha)_{\alpha\in I}$ be a trivializing cover for the bundle $\projectiongreen\piSymp:\SymFr M\to M$ with transition functions $g\AB:U_\alpha\cap U_\beta\to \Sp(2n,\RR)$.
As it is a principal fibre bundle, trivializations $\psi_\alpha:\projectiongreen{\piSymp\mo}(U_\alpha) \to U_\alpha\times \Sp(2n,\RR)$ are determined uniquely by sections $f_\alpha:U_\alpha\to \projectiongreen{\piSymp\mo}(U_\alpha)$ according to
$$
\psi_\alpha\mo(m,g) = f_\alpha(m)\cdot g
\mapob,
$$
and thus for $m\in U_\alpha\cap U_\beta$ we have
$$
f_\beta(m) = f_\alpha(m)\cdot g\AB(m)
\mapob.
$$
Now by definition $f_\alpha(m)$ is a symplectic frame at $m$ and thus determines a bijection (denoted by the same symbol) ${f_\alpha(m)}:\CanLagFr\to \SymFr_m M$ \recalt{symframegivesbijection}.
In this way we obtain a trivialization of the bundle $\projectiongreen\piLag:\LagFr M\to M$ by
\begin{moneq}[localtrivLagFrMviaSympFr]
\chi_\alpha\mo: U_\alpha\times \CanLagFr \to \projectiongreen{\piLaglos\mo}(U_\alpha)
\qquad,\qquad
\bigl(m,\begin{pmatrix} U\\ V\end{pmatrix}\bigr) \mapsto {f_\alpha(m)}\cdot\begin{pmatrix} U\\ V\end{pmatrix}
\mapob.
\end{moneq}
On $U_\alpha\cap U_\beta$ we thus have
\begin{align*}
\chi_\beta\mo\bigl(m,\begin{pmatrix} U\\ V\end{pmatrix}\bigr) 
&
= f_\beta(m)\cdot \begin{pmatrix} U\\ V\end{pmatrix}
=
\bigl(f_\alpha(m)\cdot g\AB(m)\bigr) \cdot \begin{pmatrix} U\\ V\end{pmatrix}
\\
&
=
f_\alpha(m)\cdot \bigl(g\AB(m) \cdot \begin{pmatrix} U\\ V\end{pmatrix} \bigr)
=
\chi_\alpha\mo\bigl(m, g\AB(m)\cdot \begin{pmatrix} U\\ V\end{pmatrix}  \bigr)
\mapob,
\end{align*}
where the third equality is the only one that is not obviously true (but true nevertheless).
We thus have
\begin{moneq}[transitionfunctionsLagFrM]
(\chi_\alpha\scirc \chi_\beta\mo)\bigl(m,(U,V)\bigr) = \bigl(m, g\AB(m)\cdot\begin{pmatrix} U\\ V\end{pmatrix} \bigr)
\mapob,
\end{moneq}
proving that $\LagFr M$ is indeed associated to $\SymFr M$ by the left-action of $\Sp(2n,\RR)$ on $\CanLagFr$.

To show that the right-actions of $\GlnC$ correspond, we compare \recalf{defofrightGlnCactiononLagrangians} with \recalt{symframegivesbijection}, which tells us immediately that we have the equality
\begin{moneq}
\bigl( (e;f)\cdot \begin{pmatrix} U\\ V\end{pmatrix}\bigr)\cdot A
= 
(e;f)\cdot \begin{pmatrix} UA\\ VA\end{pmatrix}
=
(e;f)\cdot (\begin{pmatrix} U\\ V\end{pmatrix}\cdot A)
\end{moneq}
for all $A\in \GlnC$ (with $(e;f)$ a symplectic frame and $(U,V)\in \CanLagFr$).
Applying this equality to the definition of the trivialization $\chi_\alpha\mo$ shows that the right-actions correspond.
\end{preuve}

\begin{definition}{Remark}
The local trivializations for $\LagFr M$ given in the proof of \recalt{LagFrInducedbySymFr} should not be confused with the local trivializations of an abstract associated bundle \recalt{constructionofassociatedbundles}. 
The general theory associates to a local trivializing section $f_\alpha$ of $\SymFr M$ a local trivialization $\Phi_\alpha$ defined by 
$$
\Phi_\alpha\mo(m,\begin{pmatrix} U\\ V\end{pmatrix})
=
\pi_\sim\bigl(f_\alpha(m),   \begin{pmatrix} U\\ V\end{pmatrix}\bigr)
\mapob,
$$
where $\pi_\sim:\SymFr M\times \CanLagFr \to B$ is the canonical projection onto the (abstract) orbit space\slash associated bundle.
The general theory then tells us that the maps $\Phi_\alpha\scirc \Phi_\beta\mo$ are exactly given by \recalf{transitionfunctionsLagFrM}.
And indeed, the bundle isomorphism mentioned at the beginning of the proof of \recalt{LagFrInducedbySymFr} identifies $\pi_\sim\bigl(f_\alpha(m),   (U,V)\bigr)$ with $f_\alpha(m)\cdot (U,V)$, and hence the abstract trivialization $\Phi_\alpha$ becomes the trivialization $\chi_\alpha$ under this identification.
One could say that the trivializations of $\LagFr M$ given in the proof of \recalt{LagFrInducedbySymFr} are concrete realizations of the abstract trivializations given by the general theory.

\end{definition}

\begin{proclaim}[CanLagFrpistypicalfiberLagFrpM]{Corollary}
The (sub)bundle $\pi_L:\LagFrp M\to M$ has $\CanLagFrp$ as typical fiber and is associated to the principal $\Sp(2n,\RR)$-bundle $\projectiongreen\piSymp:\SymFr M\to M$ by the left-action of $\Sp(2n,\RR)$ on $\CanLagFrp$. 

\end{proclaim}

\begin{proclaim}[bijectionCanLagFrpwithBtimesGlnC]{Lemma \cite[p90]{Sn80}}
Let $\Ball$ be the unit ball of symmetric matrices in $M(n,\CC)$:
$$
\Ball=\{\, W\in M(n,\CC) \mid W^t = W \ \&\ \Vert W\Vert\le 1\,\}
\mapob,
$$
where $\Vert\ \Vert$ denotes the operator norm on $M(n,\CC)$.
Then the map $\Phi:\CanLagFrp\to \Ball\times \GlnC$ defined by
$$
\Phi(U,V) = \bigl(\ (U+iV)(U-iV)\mo\ ,\ U-iV\ \bigr)
$$
is a bijection with inverse 
$$
\Phi\mo(W,C) = \bigl(\ \tfrac12\,(\oneasmat + W)C\ ,\ \tfrac i2\,(\oneasmat-W)C \ \bigr)
\mapob.
$$
Moreover, $\Phi$ is equivariant with respect to the right-actions of $\GlnC$.

\end{proclaim}

\begin{proclaim}{Lemma}
We define a left-action of $\SpnR$ on $\Ball\times \GlnC$ by
\begin{moneq}
g\cdot(W,C) = \Phi\bigl( g\cdot\Phi\mo(W,C) \bigr)
\mapob,
\end{moneq}
which thus is the left-action of $\SpnR$ on $\CanLagFrp$ transported via the bijection $\Phi$ to $\Ball\times \GlnC$.
Then there exists a left-action of the group $\SpnR$ on $\Ball$ and a map $\alpha:\SpnR\times \Ball \to \GlnC$ such that this left-action of $\SpnR$ on $\Ball\times \GlnC$ is given by
\begin{moneq}[definitionofalpha]
g\cdot (W,C)
=
(\ g\cdot W\ ,\ \alpha(g,W)\,C\ )
\mapob.
\end{moneq}

\end{proclaim}

\begin{definition}{Definition\slash Construction}
Once we dispose of the bijection $\Phi:\CanLagFrp\to \Ball\times \GlnC$, we can define the (topological) space $\CanLagFrpt$ as
\begin{moneq}
\CanLagFrpt = \Ball\times \MlnC
\end{moneq}
together with the projection $\projectiongreen \pCanLag:\CanLagFrpt\to \CanLagFrp$ given by
\begin{moneq}
\projectiongreen \pCanLag\bigl(W,(C,z)\bigr) = \Phi\mo(W,C)
\mapob,
\end{moneq}
which is a double covering map. 
Moreover, the natural free right-action of $\MlnC$ on $\CanLagFrpt$ is compatible with the $\GlnC$ action on $\CanLagFrp$ in the sens that the following diagram is commutative:
\begin{moneq}[compatibilityMlactionandGlactiononCanLagFrp]
\begin{CD}
\CanLagFrpt \times \MlnC @>>> \CanLagFrpt
\\
@V\projectiongreen {\pCanLag\times \rhoMl} VV @VV\projectiongreen {\pCanLag} V
\\
\CanLagFrp \times \GlnC @>>> \CanLagFrp\rlap{\mapob.}
\end{CD}
\end{moneq}

\end{definition}

\begin{proclaim}[compatibilityMpactionandSpactiononCanLagFr]{Lemma \cite[p92]{Sn80}}
There exists a unique map $\alphat:\MpnR\times \Ball\to \MlnC$ such that the formula
\begin{moneq}
\gt\cdot \bigl(W,(C,z)\bigr) 
=
\bigl(\ \projectiongreen\rhoMp(\gt)\cdot W\ ,\ \alphat(\gt,W)\cdot(C,z)\ \bigr)
\end{moneq}
defines a left-action of $\MpnR$ on $\Ball\times \MlnC=\CanLagFrpt$ commuting with the right action of $\MlnC$ and such that the following diagram is commutative:
$$
\begin{CD}
\Mp(2n,\RR) \times \CanLagFrpt  @>>> \CanLagFrpt
\\
@V\projectiongreen{\rhoMp\times \pCanLag} VV @VV\projectiongreen \pCanLag V
\\
\Sp(2n,\RR) \times \CanLagFrp @>>> \CanLagFrp\rlap{\mapob.}
\end{CD}
$$

\end{proclaim}

\begin{proclaim}[formofalphatinMlnC]{Corollary}
For all $\gt\in \MpnR$ and all $W\in \Ball$ we have the equality
\begin{moneq}
\alphat(\gt,W) = \bigl(\, \alpha\bigl(\projectiongreen\rhoMp(\gt),W\bigr)\, ,\, z\, \bigr) \in \MlnC
\quad\text{with}\quad
z^2 = \det\bigl(\alpha(\projectiongreen\rhoMp(\gt),W) \bigr)
\mapob.
\end{moneq}

\end{proclaim}

\begin{definition}{Summary so far}
In \recalt{CanLagFrpistypicalfiberLagFrpM} we have shown that $\CanLagFrp$ is the typical fiber of $\LagFrp M$.
Moreover, $\CanLagFrp$ has a left-action of $\SpnR$ and a (commuting) right-action of $\GlnC$, the latter being compatible with the right-action of $\GlnC$ on $\LagFrp M$.
We also have defined $\CanLagFrpt$ with its double covering map $\projectiongreen \pCanLag:\CanLagFrpt\to \CanLagFrp$. This (topological) space has a left-action of $\MpnR$ and a (commuting) right-action of $\MlnC$. 
Moreover, the projection $\projectiongreen \pCanLag$ intertwines these actions with the actions of $\SpnR$ and $\GlnC$ as shown in \recalt{compatibilityMlactionandGlactiononCanLagFrp} and \recalt{compatibilityMpactionandSpactiononCanLagFr}.
In the next section we will construct $\LagFrpt M$ as a double covering of $\LagFrp M$ with typical fiber $\CanLagFrpt$ and we will show that it does what we intended it to do: define metalinear frame bundles for all positive polarizations.

\end{definition}

\section{\texorpdfstring{$\LagFrpt M$}{LagFrptM} and induced metalinear frame bundles}
\label{sectioninducedmetalinearbundle}

Let $\SymFrt M\to M$ be a metaplectic  frame bundle.
With the preparations made in the previous section, we now  \stress{define} the bundle $\LagFrpt M$ as the (abstract) fiber bundle with typical fiber $\CanLagFrpt$ associated to the principal $\MpnR$-bundle $\SymFrt M$ and the action of $\MpnR$ on $\CanLagFrpt$.
More precisely, we define $\LagFrpt M$ as the orbit space of $\SymFrt M \times \CanLagFrpt$ under the action of $\MpnR$ given by
$$
\gt\cdot \bigl(\ft, (W,\Ct)\bigr) = \bigl(\,\ft\cdot \gt\ ,\  \gt\mo\cdot (W, \Ct)\,\bigr)
$$
and we denote $\projectiongreen\pit_\sim:\SymFrt M \times \CanLagFrpt \to \LagFrpt M$ the canonical projection.
The bundle projection $\projectiongreen\pitLag:\LagFrpt M \to M$ is the unique map making the following diagram commutative
$$
\CD
\SymFrt M \times \CanLagFrpt @>\projectiongreen\pit_\sim>> \LagFrpt M
\\
@V\projectiongreen{\pitSymp'}VV @VV\projectiongreen\pitLag V
\\
M @= M \rlap{\mapob,}
\endCD
$$
where $\projectiongreen{\pitSymp'}:\SymFrt M \times \CanLagFrpt \to M$ denotes the projection $(\ft, \varpi)\mapsto \pitSymp(\ft)$.
As $\projectiongreen\pSymp \times \projectiongreen\pCanLag$ intertwines the actions of $\MpnR$ and $\SpnR$, there exists a unique map $\projectiongreen \pLagpd : \LagFrpt M\to \LagFrp M$ such that the following diagram is commutative:
$$
\CD
\SymFrt M \times \CanLagFrpt
@>\pit_\sim>>
\LagFrpt\rlap{\quad $\cong\quad\SymFrt M \times_{\MpnR} \CanLagFrpt$}
\\
@V\projectiongreen\pSymp \times \projectiongreen\pCanLag VV @VV\projectiongreen \pLagpd V
\\
\SymFr M \times \CanLagFrp
@>>\Phi>
\LagFrp M \rlap{\quad $\cong\quad\SymFr M \times_{\SpnR} \CanLagFrp$\mapob,}
\endCD
\kern11em
$$
where $\Phi$ denotes the \myquote{projection} $\Phi(f,(U,V)) = f\cdot (U,V)$ used in the proof of \recalt{LagFrInducedbySymFr}.
And as the right-action of $\MlnC$ commutes with the action of $\MpnR$, we have an induced right-action of $\MlnC$ on $\LagFrpt M$.

\begin{proclaim}{Lemma}
The map $\projectiongreen \pLagpd:\LagFrpt M \to \LagFrp M$ is bundle map, a $2$-$1$ covering, and intertwines the free right-actions of $\MlnC$ on $\LagFrpt M$ and of $\GlnC$ on $\LagFrp M$.

\end{proclaim}

\begin{proclaim}[metalinearbundleforallpositivepolarizations]{Corollary}
Let $P$ be a positive Lagrangian distribution and $\Fr P\subset \LagFrp M$ its associated frame bundle. Then $\projectiongreen {\pLag\mo}(\Fr P) \subset \LagFrpt M$ is a metalinear frame bundle for $P$.

\end{proclaim}

With \recalt{metalinearbundleforallpositivepolarizations} we have achieved our goal: a unified way to obtain a metalinear frame bundle for all positive Lagrangian frame bundles.
However, it is not a very useful description, as it does not give us an explicit recipe how to obtain the transition functions of such a metalinear frame bundle, transition functions that obviously should depend upon the choice for the metaplectic frame bundle $\SymFrt M$.

\begin{proclaim}{A recipe for the transition functions}
Let $\SymFrt M$ be a metaplectic frame bundle, $P$ a positive Lagrangian distribution and $\Frt P$ the induced metalinear frame bundle of $P$.
Then one can obtain the transition functions of $\Frt P$ by the following procedure.
\begin{enumerate}
\item
\label{step1inducedmetalinearbundle}
Choose a nice cover $(U_\alpha)_{\alpha\in I}$ of trivializing charts for both bundles $\SymFrt M$ and $\Fr P$ simultaneously. 

\item
\label{step2inducedmetalinearbundle}
Choose (local) trivializing sections $\ft_\alpha:U_\alpha\to \SymFrt M$ and $s_\alpha:U_\alpha\to \Fr P$ for the principal fiber bundles $\SymFrt M$ and $\Fr P$ respectively with associated transition functions $\gt\AB:U_\alpha\cap U_\beta\to \MpnR$ and $N\AB:U_\alpha\cap U_\beta\to \GlnC$.

\item
\label{step3inducedmetalinearbundle}
Determine the functions $\sigma_\alpha:U_\alpha\to \CanLagFrp$ by
$$
s_\alpha(m) = f_\alpha(m)\cdot \sigma_\alpha(m)
\mapob,
$$
where $f_\alpha=\projectiongreen \pSymp\scirc \ft_\alpha$ (which are trivializing sections of $\SymFr M$.

\item
\label{step4inducedmetalinearbundle}
Choose smooth lifts\slash maps $\sigmat_\alpha:U_\alpha\to \CanLagFrpt$ such that we have $\projectiongreen \pCanLag \scirc \sigmat_\alpha = \sigma_\alpha$.

\item
\label{step5inducedmetalinearbundle}
Determine the smooth functions $W_1,W_2:U_\alpha\cap U_\beta\to \Ball$ and $\Ct_1, \Ct_2:U_\alpha\cap U_\beta\to \MlnC$ such that we have (recall the equality $\Ball\times \MlnC=\CanLagFrpt$)
\begin{moneq}
\gt\AB(m)\cdot \sigmat_\beta(m) = (W_1(m),\Ct_1(m)) 
\qquad\text{and}\qquad
\sigmat_\alpha(m) = (W_2(m),\Ct_2(m))
\mapob.
\end{moneq}

\end{enumerate}
Then the transition functions $\Nt\AB:U_\alpha\cap U_\beta\to \MlnC$ of $\Frt P$ are defined by
\begin{moneq}
\Nt\AB(m) = \Ct_2(m)\mo \cdot \Ct_1(m)
\mapob.
\end{moneq}

\end{proclaim}

\begin{preuve}
In order to justify this recipe, we have to show that all steps make sense and indeed yield the transition functions of $\Frt P$.
So we start with steps (\ref{step1inducedmetalinearbundle}) and (\ref{step2inducedmetalinearbundle}). As the bundles in question are (supposed to be) locally trivial, such a cover always exists and the transition functions associated to the local sections are defined by the equalities
$$
\ft_\beta(m) = \ft_\alpha(m) \cdot \gt\AB(m)
\qquad\text{and}\qquad
s_\beta(m) = s_\alpha(m)\cdot N\AB(m)
\mapob.
$$
Focussing for the moment on the bundle $\SymFrt M$, we note that (by definition of the projection map $\projectiongreen \pSymp:\SymFrt M\to \SymFr M$) the maps $f_\alpha=\projectiongreen \pSymp\scirc \ft_\alpha:U_\alpha\to \SymFr M$ are trivializing sections of $\SymFr M$ and that the functions $g\AB=\projectiongreen\rhoMp\scirc \gt\AB$ are the corresponding transition functions. 
Moreover, the local sections $\ft_\alpha$ define (abstract) trivializations $\chit_\alpha:\pitLag\mo(U_\alpha) \to U_\alpha \times \CanLagFrpt$ of the (associated) bundle $\LagFrpt M \to M$ by
$$
\chit_\alpha\mo\bigl(m,(W,\Ct)\bigr) = \pit_\sim\bigl(\ft_\alpha(m),(W,\Ct)\bigr)
$$
with the property that they are related according to
\begin{moneq}[transitionfunctionsLagFrpt]
(\chit_\alpha \scirc \chit_\beta\mo)\bigl(m,(W,\Ct)\bigr) = \bigl( m, \gt\AB(m)\cdot (W,\Ct)\bigr)
\mapob.
\end{moneq}
When we compare these formul{\ae} with the formul{\ae} for the (concrete) trivializations $\chi_\alpha:\piLaglos\mo(U_\alpha) \to U_\alpha \times \CanLagFrp$ of $\LagFrp M$ as given in the proof of \recalt{LagFrInducedbySymFr}, it is easy to show that the projection map $\projectiongreen \pLagpd : \LagFrpt M\to \LagFrp M$ is defined in terms of these trivializations by
\begin{align}
\chi_\alpha \scirc \projectiongreen \pLagpd \scirc \chit_\alpha\mo: U_\alpha\times \CanLagFrpt 
&
\to U_\alpha \times \CanLagFrp
\notag
\\ 
\bigl(\,m,(W,\Ct)\,) 
&
\mapsto \bigl(m,\projectiongreen \pCanLag (W,\Ct)\,\bigr)
\mapob.
\label{projectionpsubLintrivializations}
\end{align}

As $\Fr P$ is (can be seen as) a subbundle of $\LagFrp M$ ($P$ is supposed to be positive), the $s_\alpha$ are also sections of $\LagFrp M$ and thus there exist smooth functions $\sigma_\alpha:U_\alpha\to \CanLagFrp$ defined by
\begin{moneq}
\chi_\alpha\bigl(s_\alpha(m)\bigr) = \bigl( m, \sigma_\alpha(m) \bigr)
\qquad\Longleftrightarrow\qquad
s_\alpha(m) = f_\alpha(m)\cdot \sigma_\alpha(m)
\mapob,
\end{moneq}
showing that step (\ref{step3inducedmetalinearbundle}) is well defined.
It follows that $\Fr P$ as subbundle of $\LagFrp M$ is given in terms of the trivializations by
\begin{align*}
\chi_\alpha\bigl( \projectiongreen{\piLaglos\mo}(U_\alpha) \cap \Fr P\bigr) 
&
= 
\Bigl\{\,\bigl(m,\sigma_\alpha(m)\cdot N\bigr) \mid m\in U_\alpha \ ,\ N\in \GlnC\,\Bigr\}
\subset
U_\alpha\times \CanLagFrp
\mapob.
\end{align*}
From this it follows immediately that the inverse image $\projectiongreen {\pLag\mo}(\Fr P)\subset \LagFrpt M$ is given in terms of the trivializations by (use \recalf{projectionpsubLintrivializations})
\begin{align*}
&
\chit_\alpha\bigl( \projectiongreen{\pitLag\mo}(U_\alpha) \cap \projectiongreen{ \pLag\mo}(\Fr P)\bigr) 
= 
\\
&\kern3em
\Bigl\{\,\bigl(m,\bigl(W,\Ct)\bigr) \ \bigm\vert\  m\in U_\alpha \ ,\  \exists N\in \GlnC: 
\\
&\kern14em
\projectiongreen \pCanLag(W,\Ct) = \sigma_\alpha(m)\cdot N\,\Bigr\}
\subset
U_\alpha\times \CanLagFrpt
\mapob.
\end{align*}

Now the sets $U_\alpha$ are contractible (by definition of a nice cover), so there exist lifts of the (smooth) maps $\sigma_\alpha:U_\alpha\to \CanLagFrp$ to smooth maps $\sigmat_\alpha:U_\alpha\to \CanLagFrpt$, \ie, we have a commutative diagram
$$
\begin{matrix}
&&\CanLagFrpt
\\
\noalign{\vskip2\jot}
&\llap{\raise5pt\hbox{$\scriptstyle\sigmat_\alpha$}}\nearrow & \downarrow \rlap{$\scriptstyle\projectiongreen \pCanLag$}
\\
\noalign{\vskip2\jot}
U_\alpha & \underset{\sigma_\alpha}{\to} & \CanLagFrp\rlap{\mapob,}
\end{matrix}
$$
justifying step (\ref{step4inducedmetalinearbundle}).
It follows immediately that we have
\begin{align*}
\chit_\alpha\bigl( \projectiongreen{\pitLag\mo}(U_\alpha) \cap \projectiongreen{ \pLag\mo}(\Fr P)\bigr) 
&
= 
\bigl\{\,\bigl(m,\sigmat_\alpha(m)\cdot \Ct\bigr) \mid m\in U_\alpha \ ,\ \Ct\in \MlnC\,\bigr\}
\end{align*}
and that the maps $\st_\alpha:U_\alpha\to \LagFrpt$ defined by
\begin{moneq}
\st_\alpha(m) = \chit_\alpha\mo\bigl( m,\sigmat_\alpha(m) \bigr)
\end{moneq}
are trivializing sections for $\projectiongreen{ \pLag\mo}(\Fr P)\subset \LagFrpt M$ satisfying $\projectiongreen\pLagpd \scirc \st_\alpha = s_\alpha$ (use \recalf{projectionpsubLintrivializations}).

We now compute, for $m\in U_\alpha\cap U_\beta$:
\begin{align*}
\bigl(m, g\AB(m)\cdot \sigma_\beta(m) \bigr)
\kern0.5em
&
\oversetalign{\text{\recalt{transitionfunctionsLagFrM}}}\to=
\kern0.5em
(\chi_\alpha\scirc \chi_\beta\mo)\bigl(m, \sigma_\beta(m)\bigr)
=
(\chi_\alpha\scirc \chi_\beta\mo)\bigl(\chi_\beta\bigl( s_\beta(m)\bigr)\bigr)
\\&
=
\chi_\alpha\bigl( s_\beta(m)\bigr)
=
\chi_\alpha\bigl( s_\alpha(m)\cdot N\AB(m)\bigr)
\\&
=
\bigl(m, \sigma_\alpha(m)\cdot N\AB(m)\bigr) 
\end{align*}
where for the last equality we used that the trivializations are compatible with the right-actions of $\GlnC$ \recalt{LagFrInducedbySymFr}.
And thus we have the equality
\begin{moneq}[sigmabisgbasigmaaAab]
g\AB(m)\cdot \sigma_\beta(m) = \sigma_\alpha(m)\cdot N\AB(m) 
\mapob.
\end{moneq}
Using the fact that $\pCanLag:\CanLagFrpt \to \CanLagFrp$ is a $2$-$1$ covering and that the action of $\MlnC$ on $\CanLagFrpt$ is free, it follows easily that there exist unique functions $\Nt\AB:U_\alpha\cap U_\beta\to \MlnC$ such that we have:
\begin{moneq}[liftedequalitysigmatildes]
\gt\AB(m)\cdot \sigmat_\beta(m) = \sigmat_\alpha(m)\cdot \Nt\AB(m) 
\mapob.
\end{moneq}
Using this equality, we compute, again for $m\in U_\alpha\cap U_\beta$:
\begin{align*}
\st_\beta(m)
&
=
\chit_\beta\mo\bigl(m,\sigmat_\beta(m)\bigr)
=
\chit_\alpha\mo\bigl((\chit_\alpha\scirc\chit_\beta\mo)\bigl(m,\sigmat_\beta(m)\bigr)\bigr)
\\&
\oversetalign{\recalf{transitionfunctionsLagFrpt}}\to=
\kern0.5em
\chit_\alpha\mo\bigl(m,\gt\AB(m)\cdot \sigmat_\beta(m)   \bigr)
=
\chit_\alpha\mo\bigl(m,\sigmat_\alpha(m)\cdot \Nt\AB(m)  \bigr)
\\&
=
\st_\alpha(m)\cdot \Nt\AB(m)
\mapob,
\end{align*}
showing that the $\Nt\AB$ are the transition functions of the bundle $\projectiongreen{ \pLag\mo}(\Fr P)$.
And as we also have $\rhoMl\bigl(\Nt\AB(m)\bigr) = N\AB(m)$, we thus have shown that $\projectiongreen{ \pLag\mo}(\Fr P)$ indeed is a metalinear frame bundle for $P$.

In order to get an explicit expression for the transition functions $\Nt\AB$ of this metalinear frame bundle (and to show that they are indeed smooth), we follow step (\ref{step5inducedmetalinearbundle}) and compute the smooth functions $W_i$ and $\Ct_i$ as indicated.
According to \recalf{liftedequalitysigmatildes} we thus have the equality
\begin{align*}
\bigl(W_1(m),\Ct_1(m)\bigr)
&
=
\gt\AB(m)\cdot \sigmat_\beta(m) 
=
\sigmat_\alpha(m) \cdot \Nt\AB(m) 
\\&
= 
\bigl(W_2(m),\Ct_2(m)\bigr)\cdot \Nt\AB(m)
= 
\bigl(W_2(m),\Ct_2(m)\cdot \Nt\AB(m)\bigr)
\mapob,
\end{align*}
and in particular $\Ct_1(m)=\Ct_2(m)\cdot \Nt\AB(m)$.
The final result follows immediately.
\end{preuve}

\section{The compatibility condition}

Once we have defined the metalinear frame bundles induced by the metaplectic frame bundle, we have to attack the question whether these metalinear frame bundles are compatible for two compatible positive Lagrangian distributions $P_1$ and $P_2$, \ie, whether there exists a global function $\deltakt$ on $\PairPolPt{12} M$.
The search for a \myquote{globally defined} function $\deltakt$ (for all positive Lagrangian distributions at the same time) is slightly more subtle than the naive approach suggests.
The reason is that the bundle $\PairPolPt{12} M$ depends upon the (real) intersection foliation $D$ defined by $D^\CC = \Pb_1\cap P_2$ (and thus in particular on its dimension $k$) via the fact that we use special frames for $P_1$ and $P_2$ whose first $k$ vectors are real and coincide (and thus form a basis of $D$).
At first sight it thus seems natural to define a subbundle of $\LagFrp M\times \LagFrp M$ as those pairs $(u,v)$ of positive Lagrangian frames whose first $k$ vectors are real and coincide.
However, there seems to be no easy way to describe the typical fiber of such a bundle.
And without such a description, the quest for a lift to metaframes of a generalized function $\deltakt$ seems to be hopeless.
The approach we will take is more restrictive and consists of fixing not only the dimension $k$, but the (isotropic) distribution $D$ itself. 
We thus will look at pairs of (positive) Lagrangian frames whose first $k$ vectors are real, coincide and form a frame\slash basis of $D$.
For this more restricted subbundle we can find a nice description of the typical fiber, a description that will allow us to define our generalization of the function $\deltakt$.

\begin{definition}{Definitions}
Let $D\subset TM$ be an isotropic distribution of dimension $k$ on $M$.
We define the subbundles $\PairLagFrpynD[D] M\subset \LagFrpyn M\times_M \LagFrpyn M$ (everywhere or nowhere the subscript $+$) by
\begin{align*}
\PairLagFrpynD[D]\caprestricted_m M
&
=
\bigl\{\ (u,v)\in \LagFrpyn\caprestricted_m M \times \LagFrpyn\caprestricted_m M \mid
\\
&\kern7em
\forall 1\le i\le k : 
\ub_i = u_i = v_i
\ \&\ 
(u_1, \dots,u_k)\in \Fr D_m
\ \bigr\}
\mapob,
\end{align*}
\ie, those pairs of Lagrangian frames whose first $k$ elements coincide and form a basis of $D_m$ (and that are positive in case of the subscript $+$).
This bundle has a natural right-action of $\Glkd$.

On $\PairLagFrpynD[D]M$ we define the function $\delta_D:\PairLagFrpynD[D]M \to \CC$ by
\begin{moneq}[defofdeltasubD]
\delta_D(u,v)
=
\det\bigl( -i\cdot\omega(\ub_{i}, v_{j})_{ij,=k+1}^n \bigr)
\mapob.
\end{moneq}
We also define the subbundle $\PairLagFrpDt[D]M\subset \LagFrpt M\times_M \LagFrpt M$ by
\begin{align*}
\PairLagFrpDt[D]M\caprestricted_m 
&
=
\{\ (\ut,\vt)\in \LagFrpt\caprestricted_m M \times \LagFrpt\caprestricted_m M \mid
\bigl( \projectiongreen \pLagpd(\ut), \projectiongreen \pLagpd(\vt) \bigr) \in \PairLagFrpD[D]\caprestricted_m M
\ \}
\mapob,
\end{align*}
together with the projection $\projectiongreen \pPairLag:\PairLagFrpDt[D]M \to \PairLagFrpD[D] M$ defined as 
$$
\projectiongreen \pPairLag(\ut,\vt) = \bigl(\projectiongreen\pLagpd(\ut), \projectiongreen\pLagpd(\vt) \bigr)
\mapob.
$$

\end{definition}

\begin{proclaim}{Lemma}
For any $(u,v)\in \PairLagFrpynD[D]M$ and any $(g_1,g_2)\in \Glkd$ of the form \recalf{specialformGkld} we have
\begin{align*}
\delta_D\bigl( (u,v)\cdot (g_1,g_2)\bigr)
&
=
\delta_D(u,v) \cdot \overline{\det(D_1)}\cdot{\det(D_2)}
\notag
\\
&
=
\delta_D(u,v) \cdot \overline{\det(g_1)}\cdot{\det(g_2)} \cdot \det(A)^{-2}
\mapob.
\end{align*}

\end{proclaim}

\begin{proclaim}[restrictionofdeltasubDgivesdeltak]{Lemma}
Let $P_1$ and $P_2$ be two compatible Lagrangian distributions with $\Pb_1\cap P_2=D^\CC$. 
Then $\PairPolP{12}$ is a subbundle of $\PairLagFrD[D] M$ and the restriction of $\delta_D$ to this subbundle is the function $\deltak$ defined in \recalf{definitionofdelta2}.
Moreover, when the $P_i$ are positive, then $\PairPolP{12}$ is a subbundle of $\PairLagFrpD[D] M \subset \PairLagFrD[D] M$

\end{proclaim}

\begin{proclaim}[PairPolPtsubbundlePairLagFrpDt]{Lemma}
Let $P_1$ and $P_2$ be two compatible positive Lagrangian distributions with $\Pb_1\cap P_2=D^\CC$. 
If we take $\Frt{P_i} = \projectiongreen{\pLag\mo}(\Fr{P_i})$ as metalinear frame bundles for them, then the bundle $\PairPolPt{12}\to M$ is a subbundle of $\PairLagFrpDt[D]M$.

\end{proclaim}

With these preparations, it now \myquote{suffices} to define a function $\deltat_D:\PairLagFrpDt[D]M\to\CC$ such that the restriction to $\PairPolPt{12} \subset \PairLagFrpDt[D]M$ is the sought for function $\deltakt$.
To do so, we need the typical fibre of $\PairLagFrpD[D]M$ in terms of the typical fibre $\CanLagFrp$ of $\LagFrp M$ and we have to describe the function $\delta_D$ in terms of this typical fibre.
Now the identification between a fiber of $\LagFrp M$ and $\CanLagFrp$ is given by a symplectic frame $(e;f)$, but such an identification does not take into account that the first $k$ vectors of our frames for $\PairLagFrpD[D]M$ form a frame of $D$. 
And without that information, it is hard to describe the typical fiber of $\PairLagFrpDt[D]M$ as a subset of $\CanLagFrp \times \CanLagFrp$.

The idea thus is to define $D$-adapted symplectic frames that include this information. 
In terms of $D$-adapted symplectic frames, the description of our Lagrangian frames whose first $k$ vectors form a basis of $D$ becomes \myquote{nice} \recalt{appearanceofreducesCanLagFr}. 
But then we have to show that we can indeed do so everywhere, which we do by showing that the subbundle of $D$-adapted symplectic frames is (again) a principal fiber bundle and that all for us relevant fiber bundles are associated bundles to this principal one or to its \myquote{lift} to metaplectic frames.

\begin{definition}{Definitions}
Let $D\subset TM$ be an isotropic (real) distribution of dimension $k$, with its associated coisotropic distribution $E=D^\perp\supset D$, the symplectic orthogonal of $D$ of dimension $2n-k$.
We will say that a symplectic frame $(e;f)$ of $T_mM$ is \stresd{$D$-adapted} if it satisfies the conditions
\begin{enumerate}
\item
$e_1, \dots, e_k$ is a basis for $D_m$ and

\item
$e_1, \dots, e_n, f_{k+1}, \dots, f_n$ is a basis for $E_m$.

\end{enumerate}
We define the subbundle $\SymFr_D M\subset \SymFr M$ as the subbundle of $D$-adapted symplectic frames:
\begin{moneq}
\SymFr_D M = 
\{\, (e;f)\in \SymFr M \mid (e;f) \text{ is $D$-adapted}\,\}
\mapob.
\end{moneq}
Associated to this subbundle we have the subbundle $\SymFrtD M = \projectiongreen{ \pSymplos\mo}(\SymFr_DM)\subset \SymFrt M$ of the metaplectic frame bundle defined as the inverse image of $\SymFr_DM$.

A local section $\psi:U\to \SymFr M$ is called \stresd{$D$-adapted} if for all $m\in U$ the symplectic frame $\psi(m)$ is $D$-adapted.
And by a \stresd{$D$-adapted trivializing cover (for $\SymFr M$)} we will mean a cover $(U_\alpha)_{\alpha\in I}$ and local (trivializing) sections $\psi_\alpha:U_\alpha\to \SymFr M$ that are $D$-adapted.

We define the subgroup $\SpknR\subset \Sp(2n,\RR)$ as consisting of the elements $g\in \Sp(2n,\RR)$ of the form
\begin{moneq}[defofsubgroupSpk2nR]
g = 
\begin{pmatrix} 
A_g^t & B_g & C_g & D_g
\\
\mathbf0 & T_{1r} & E_g & T_{2r}
\\
\mathbf0 & \mathbf0 & A_g\mo & \mathbf0
\\
\mathbf0 & T_{3r} & F_g & T_{4r}
\end{pmatrix}
\mapob,
\end{moneq}
with $A_g\in \Gl(k,\RR)$, $B_g,C_g,D_g,E_g,F_g$ arbitrary real matrices of the appropriate sizes and
$$
g_r
=
\begin{pmatrix}
T_{1r} & T_{2r} \\ T_{3r} & T_{4r} 
\end{pmatrix}
\in \Sp\bigl(2(n-k),\RR\bigr)
\mapob.
$$
Associated to $\SpknR$ we define the subgroup $\MpknR\subset \Mp(2n,\RR)$ as the inverse image of $\SpknR$ under the projection $\projectiongreen\rhoMp:\Mp(2n,\RR)\to\Sp(2n,\RR)$.

\end{definition}

\begin{proclaim}{Lemma \cite[p97]{Sn80}}
The bundle $\SymFr_D M$ is a principal $\SpknR$ bundle over $M$ and $\SymFrtD M$ is a principal $\MpknR$ bundle over $M$.

\end{proclaim}

\begin{proclaim}[LagFrMhassmallerstructuregroup]{Corollary}
The bundle $\LagFrpyn M\to M$ is associated to the principal $\SpknR$-bundle $\SymFr_D M\to M$ by the left-action of $\SpknR$ on $\CanLagFrpyn$.

\end{proclaim}

\begin{proclaim}[LagFrtMhassmallerstructuregroup]{Corollary}
The bundle $\LagFrpt M\to M$ is associated to the principal $\MpknR$-bundle $\SymFrtD M$ by the left-action of $\MpknR$ on $\CanLagFrpt$.

\end{proclaim}

\begin{definition}[remarkreducingstructuregroup]{Remark}
\recalt{LagFrMhassmallerstructuregroup} can be interpreted as saying that we reduce the structure group of the bundle $\LagFrpyn M$ from $\SpnR$ \recalt{LagFrInducedbySymFr} to $\SpknR$ (a similar interpretation holds for \recalt{LagFrtMhassmallerstructuregroup}).
We can also interpret it as saying that we can trivialize $\LagFrpyn M$ using only $D$-adapted symplectic frames (see \recalt{symframegivesbijection}).

\end{definition}

\begin{proclaim}[appearanceofreducesCanLagFr]{Lemma}
Let $P$ be a Lagrangian distribution satisfying $D^\CC\subset P$, let $u\in \Fr P_m$ be a frame whose first $k$ vectors belong to $D$ and let $(e;f)$ be a $D$-adapted symplectic frame. 
If we define (see \recalt{symframegivesbijection}) the matrices $U,V\in M(n,\CC)$ by
\begin{moneq}
u=(e;f)\cdot \begin{pmatrix} U\\ V\end{pmatrix}
\mapob,
\end{moneq}
then they are necessarily of the form
\begin{moneq}[specialreducedformUV]
U = \begin{pmatrix} A & B \\ \mathbf0 & U_r \end{pmatrix}
\quad\text{and}\quad
V = \begin{pmatrix} \mathbf0 & \mathbf0 \\ \mathbf0 & V_r \end{pmatrix}
\mapob,
\end{moneq}
with $A\in \Gl(k,\RR)$, the other matrices complex of the appropriate size.
Moreover, the matrices $U_r,V_r\in M(n-k,\CC)$ satisfy the conditions
\begin{enumerate}
\item
$\det(U_r^\dagger \,U_r+V_r^\dagger \,V_r)\neq0$,

\item
$U_r^t\, V_r = V_r^t\, U_r$,

\end{enumerate}
Additionally, the Lagrangian distribution $P$ is positive if and only if the matrices $U_r,V_r$ satisfy the additional condition
\begin{enumerate}
\setcounter{enumi}{2}
\item
$i\,(V_r^\dagger\, U_r - U_r^\dagger\, V_r)$ is non-negative definite.

\end{enumerate}

\end{proclaim}

\begin{definition}{A change of notation}
In \recalt{appearanceofreducesCanLagFr} we see that the conditions to belong to $\CanLagFr$ or $\CanLagFrp$ appear relatively naturally for matrices of a smaller size.
This motivates us to add a dimension indicator to the space $\CanLagFr$ and its \myquote{derived} spaces $\CanLagFrp$ and $\CanLagFrpt$.
We will do this by adding ${(n)}$ as superscript (with $n$ the dimension), \ie, we will denote these spaces now as $\CanLagFr[n]$, $\CanLagFrp[n]$ and $\CanLagFrpt[n]$.
We also add this superscript to the set $\Ball$, the bijection $\Phi$ used in \recalt{bijectionCanLagFrpwithBtimesGlnC} and the map $\alpha$ \recalf{definitionofalpha}, thus writing $\Phi^{(n)}:\CanLagFrp[n]\to \Ball^{(n)}\times \GlnC$ and $\alpha^{(n)}$.

\end{definition}

\begin{definition}[defofPairCanLagFrpyn]{Definition}
We define the subset $\PairCanLagFrpyn[n\vert k] \subset \CanLagFrpyn[n] \times \CanLagFrpyn[n]$ as (everywhere or nowhere a subscript $+$) 
\begin{align*}
\PairCanLagFrpyn[n\vert k]
&
=
\left\{\ 
\phantom{\llap{$\displaystyle\begin{pmatrix} A & B_i \\ \mathbf0 & U_{ir} \end{pmatrix}$}}
\bigl((U_1,V_1),(U_2,V_2)\bigr)\in \bigl(\CanLagFrpyn[n]\bigr)^2 \Bigm\vert 
\exists A\in \Gl(k,\RR)\ ,\ \exists B_i\ ,\ 
\right.
\\
&\kern4em
\left.
\exists (U_{ir},V_{ir})\in \CanLagFrpyn[n-k] :
U_i = \begin{pmatrix} A & B_i \\ \mathbf0 & U_{ir} \end{pmatrix}
\ \&\ 
V_i = \begin{pmatrix} \mathbf0 & \mathbf0 \\ \mathbf0 & V_{ir} \end{pmatrix}
\ \right\}
\mapob.
\end{align*}

\end{definition}

\begin{proclaim}{Lemma}
$\PairCanLagFr[n\vert k]$ and $\PairCanLagFrp[n\vert k]$ are invariant under the (diagonal) left-action of $\SpknR$ and under the right-action of $\Glkd$.

\end{proclaim}

\begin{proclaim}[LCanLagFristypicalfibrePairLagFr]{Lemma}
The bundle $\PairLagFrpynD[D] M\to M$ has $\PairCanLagFrpyn[n\vert k]$ as typical fiber and is associated to the principal $\SpknR$-bundle $\SymFr_D M\to M$ by the (diagonal) left-action of $\SpknR$ on $\PairCanLagFrpyn[n\vert k]$.
Moreover, the right-action of $\Glkd$ on $\PairLagFrpynD[D] M$ corresponds to the right-action of $\Glkd$ on $\PairCanLagFrpyn[n\vert k]$.

\end{proclaim}

\begin{proclaim}[descriptionchangefromUVtoWC]{Lemma}
For $g\in \SpknR$ of the form \recalf{defofsubgroupSpk2nR} and $(U,V)\in \CanLagFr$ of the form \recalf{specialreducedformUV}, we have
\begin{moneq}
g\cdot (U,V)
=
(\ 
\begin{pmatrix} A_g^t\, A & * \\ \mathbf0 & U_r' \end{pmatrix}
\ ,\ 
\begin{pmatrix} \mathbf0 & \mathbf0 \\ \mathbf0 & V_r' \end{pmatrix}
\ )
\quad\text{with}\quad
\begin{pmatrix} U_r' \\ V_r' \end{pmatrix}
= 
g_r\cdot 
\begin{pmatrix} U_r \\ V_r \end{pmatrix}
\mapob.
\end{moneq}
Moreover, if $(U,V)\in \CanLagFrp$, then we also have $\Phi^{(n)}(U,V) = (W,C)$ with
\begin{moneq}
W = \begin{pmatrix} \oneasmat & \mathbf0 \\ \mathbf0 & W_{r} \end{pmatrix}
\quad\text{,}\quad
C = \begin{pmatrix} A & B \\ \mathbf0 & C_{r} \end{pmatrix}
\quad\text{and}\quad
\Phi^{(n-k)}(U_r,V_r) = (W_r,C_r)
\mapob.
\end{moneq}
as well as (see \recalf{definitionofalpha})
\begin{moneq}[lemmaonreducedformsandactions]
g\cdot W = \begin{pmatrix} \oneasmat & \mathbf0 \\ \mathbf0 & g_r\cdot W_{r} \end{pmatrix}
\quad\text{and}\quad
\alpha^{(n)}(g,W) = \begin{pmatrix} A_g^t & * \\ \mathbf0 & \alpha^{(n-k)}(g_r,W_r) \end{pmatrix}
\mapob.
\end{moneq}

\end{proclaim}

\begin{definition}{Definition}
We define the subset $\PairCanLagFrpt[n\vert k] \subset \CanLagFrpt[n] \times \CanLagFrpt[n]$ as  
\begin{align}
\PairCanLagFrpt[n\vert k]
&
=
\Bigl\{\ 
\bigl(\,\bigl(W_1,(C_1,z_1)\bigr),\bigl(W_2,(C_2,z_2)\bigr)\,\bigr)\in \Migl(\,\CanLagFrpt[n]\,\Migr)^2 
\bigm\vert 
\notag
\\
&\kern12em
\bigl(\,\projectiongreen\pCanLag(W_1,C_1),\projectiongreen\pCanLag(W_2,C_2)\,\bigr)\in \PairCanLagFrp[n\vert k]
\ \Bigr\}
\notag
\\
{\text{\small(use \recalt{descriptionchangefromUVtoWC})}}\quad
&
=
\left\{\ 
\phantom{\llap{$\displaystyle\begin{pmatrix} A & B_i \\ \mathbf0 & U_{ir} \end{pmatrix}$}}
\bigl(\,\bigl(W_1,(C_1,z_1)\bigr),\bigl(W_2,(C_2,z_2)\bigr)\,\bigr)\in \Migl(\,\CanLagFrpt[n]\,\Migr)^2 
\bigm\vert 
\right.
\notag
\\
&\kern4em
\exists A\in \Gl(k,\RR)\ \exists B_i \ 
\exists (W_{ir},C_{ir})\in \Ball^{(n-k)}\times \Gl(n-k,\CC) :
\notag
\\
&\kern13em
\left.
W_i = \begin{pmatrix} \oneasmat & \mathbf0 \\ \mathbf0 & W_{ir} \end{pmatrix}
\ \&\ 
C_i = \begin{pmatrix} A & B_i \\ \mathbf0 & C_{ir} \end{pmatrix}
\ \right\}
\mapob.
\label{seconddescriptionPairCanLagFrpt}
\end{align}
Associated to $\PairCanLagFrpt[n\vert k]$ we define the projection $\projectiongreen\pPairCanLag:\PairCanLagFrpt[n\vert k]\to \PairCanLagFrp[n\vert k]$ by
\begin{moneq}
\projectiongreen\pPairCanLag\bigl(\,(W_1,C_1),(W_2,C_2)\,\bigr) = \bigl(\,\projectiongreen \pCanLag(W_1,C_1),\projectiongreen \pCanLag(W_2,C_2)\,\bigr)
\mapob.
\end{moneq}

\end{definition}

\begin{proclaim}[PairCanLagFrtinvariantunderMpk]{Lemma}
The set $\PairCanLagFrpt[n\vert k]$ is invariant under the (diagonal) left-action of $\MpknR$ and under the right-action of $\Mlkd$.

\end{proclaim}

\begin{proclaim}[LCanLagFrptistypicalfibrePairLagFr]{Lemma}
The bundle $\PairLagFrpDt[D] M\to M$ has $\PairCanLagFrpt[n\vert k]$ as typical fiber and is associated to the principal $\MpknR$-bundle $\SymFrtD M\to M$ by the (diagonal) left-action of $\MpknR$ on $\PairCanLagFrpt[n\vert k]$.
Moreover, the right-action of $\Mlkd$ on $\PairLagFrpDt[D] M$ corresponds to the right-action of $\Mlkd$ on $\PairCanLagFrpt[n\vert k]$.

\end{proclaim}

Now that we have a good description of the typical fibers of the bundles $\PairLagFrpD[D]M$ and $\PairLagFrpDt[D]M$ and that we know that these bundles are associated bundles to the (reduced) symplectic frame bundles $\SymFr_D M$ and $\SymFrtD M$, we can attack the question how to define the lift of the function $\delta_D$ to $\PairLagFrpDt[D]M$.

\begin{proclaim}[deltakintermsofCanLagFr]{Lemma}
Let $(u_1,u_2)\in \PairLagFrD[D] M\caprestricted_m$ be arbitrary and let $(e;f)$ be a $D$-adapted symplectic frame.
If we define (see \recalt{symframegivesbijection}) the matrices $U_i,V_i\in M(n,\CC)$ by
\begin{moneq}
u_i=(e;f)\cdot \begin{pmatrix} U_i\\ V_i\end{pmatrix}
\mapob,
\end{moneq}
then we have the equality (use \recalt{appearanceofreducesCanLagFr})\begin{moneq}
\delta_D(u_1,u_2)
=
\det\bigl(\,i\cdot( V_{1r}^{\dagger}\,U_{2r} - U_{1r}^{\dagger}\, V_{2r})  \,\bigr)
\mapob.
\end{moneq}

\end{proclaim}

\begin{definition}{Definition}
We define the function $\deltaLk : \PairCanLagFr[n\vert k] \to \CC$ by
\begin{moneq}
\deltaLk\bigl((U_1,V_1),(U_2,V_2)\bigr) = 
\det\bigl(\,i\cdot( V_{1r}^{\dagger}\,U_{2r} - U_{1r}^{\dagger}\, V_{2r})  \,\bigr)
\mapob.
\end{moneq}

\end{definition}

\begin{proclaim}[invarianceofdeltaLkandtransformation]{Lemma}
The function $\deltaLk$ is invariant under the (diagonal) left-action of $\SpknR$. Moreover, for all $X\in \PairCanLagFr[n\vert k]$ and all $(g_1,g_2)\in \Glkd$ of the form \recalf{specialformGkld} we have (to be compared with \recalf{tranformationpropdeltak})
\begin{align*}
\deltaLk\bigl( X\cdot (g_1,g_2)\bigr)
&
=
\deltaLk(X) \cdot \overline{\det(D_1)}\cdot{\det(D_2)}
\\
&
=
\deltaLk(X) \cdot \overline{\det(g_1)}\cdot{\det(g_2)} \cdot \det(A)^{-2}
\mapob.
\end{align*}

\end{proclaim}

\begin{proclaim}[localexpressiondeltasubD]{Lemma}
Let $\psi_\alpha:U_\alpha\to \SymFr_D M$ be a trivializing section, \ie, $\psi_\alpha(m)$ is a $D$-adapted frame for all $m\in U_\alpha$, and let $\Xi_\alpha: \projectiongreen{ \bigl(\piPairLag\bigr)\mo}(U_\alpha) \to U_\alpha \times \PairCanLagFr[n\vert k]$ be the corresponding local trivialization of $\PairLagFrD[D] M$, which thus is given by (see also \recalf{localtrivLagFrMviaSympFr})
$$
(\Xi_\alpha)\mo(m,X) = \psi_\alpha(m)\cdot X
\mapob.
$$
Then we have the equality
\begin{moneq}
\delta_D\bigl(\Xi_\alpha\mo(m,X)\bigr)
=
\deltaLk(X)
\mapob.
\end{moneq}

\end{proclaim}

\begin{proclaim}[explicitdescriptiondeltaLk]{Lemma}
Describing an element $\bigl((U_1,V_1),(U_2,V_2)\bigr)\in \PairCanLagFrp[n\vert k]$ in terms of a couple $\bigl((W_1,C_1),(W_2,C_2)\bigr)\in \bigl(\Ball^{(n)}\times \GlnC\bigr)^2$ using \recalt{descriptionchangefromUVtoWC}, the function $\deltaLk:\PairCanLagFrp[n\vert k] \to \CC$ is given by
\begin{align*}
\deltaLk\bigl(\, (\Phi^{(n)})\mo(W_1,C_1)
&
\,,\, (\Phi^{(n)})\mo(W_2,C_2) \,\bigr)
\\&
=
\det(C_{1r}^\dagger) \cdot \det\bigl(\tfrac12(\oneasmat - W_{2r}^\dagger\, W_{1r})\bigr)\cdot \det(C_{2r})
\\
&
=
\overline{\det(C_{1})} \cdot {\det(C_{2})} \cdot \det(A)^{-2} \cdot \det\bigl(\tfrac12(\oneasmat - W_{2r}^\dagger\, W_{1r})\bigr)
\mapob.
\end{align*}

\end{proclaim}

\begin{proclaim}[GammaissquarerootofdetoneminusWW]{Lemma \cite[p95]{Sn80}}
For all $n$ there exists a unique continuous function $\Gamma^{(n)}:\Ball^{(n)}\times \Ball^{(n)} \to \CC$ such that\,\footnote{Actually, the factor $\tfrac12$ inside the determinant is absent in \cite{Sn80}, but that changes the function $\Gamma^{(n)}$ only by a factor $2^n$, which does not affect the statement.} 
\begin{moneq}
\bigl(\Gamma^{(n)}(W_1,W_2)\bigr)^2 = \det\bigl(\tfrac12(\oneasmat-W_2^\dagger\, W_1)\bigr)
\qquad\text{and}\qquad
\Gamma^{(n)}(\mathbf0,\mathbf0) = 1
\mapob.
\end{moneq}

\end{proclaim}

\begin{definition}{Definition}
Inspired by \recalt{explicitdescriptiondeltaLk} we define the function $\deltaLkt:\PairCanLagFrpt[n\vert k] \to \CC$ by
\begin{moneq}
\deltaLkt\migl(\,\bigl(W_1,(C_1,z_1)\bigr),\bigl(W_2,(C_2,z_2)\bigr)\,\migr)
=
\zb_1\cdot z_2\cdot \vert\det(A)\vert\mo \cdot \Gamma^{(n-k)}(W_{1r},W_{2r})
\mapob,
\end{moneq}
where $A$ and $W_{ir}$ are as in \recalf{seconddescriptionPairCanLagFrpt}.

\end{definition}

\begin{proclaim}{Lemma}
The function $\deltaLkt$ is invariant under the (diagonal) left-action of $\MpknR$. Moreover, for all $X\in \PairCanLagFrpt[n\vert k]$ and all $(\gt_1,\gt_2)\in\Mlkd$ of the form \recalf{structuregroupPairPoltNEW} we have the equalities (to be compared with \recalf{deltaktrootofdeltak} and \recalf{behaviourdeltaktunderMlkd})
\begin{align*}
\bigl(\deltaLkt(X)\bigr)^2 
&
= \deltaLk\bigl(\projectiongreen\pPairCanLag(X)\bigr)
\\
\deltaLkt\bigl(X\cdot(\gt_1,\gt_2)\bigr) 
&
= 
\deltaLkt(X) \cdot \zb_1\cdot {z_2}\cdot \vert\det(A)\vert\mo
\mapob.
\end{align*}
(Nota Bene: here the $z_i$ and $A$ are part of the $\gt_i$, not of $X$.)

\end{proclaim}

\begin{preuve}
The two equalities are obtained by a direct computation, using \recalt{explicitdescriptiondeltaLk} and \recalt{GammaissquarerootofdetoneminusWW} for the first.
To prove that it is invariant under the left-action of $\MpknR$ we first invoke \recalt{compatibilityMpactionandSpactiononCanLagFr} and \recalt{invarianceofdeltaLkandtransformation} to show that for $\gt\in \MpknR$ and $X\in \PairCanLagFrpt[n\vert k]$ we have
\begin{moneq}
\bigl(\deltaLkt(\gt\cdot X)\bigr)^2
= 
\deltaLk\bigl(\projectiongreen\pPairCanLag(\gt\cdot X)\bigr) 
= 
\deltaLk\bigl(\projectiongreen\rhoMp(\gt)\cdot \projectiongreen\pPairCanLag(X)\bigr)
=
\deltaLk\bigl(\projectiongreen\pPairCanLag(X)\bigr)
=
\bigl(\deltaLkt(X)\bigr)^2
\mapob.
\end{moneq}
By a continuity argument it follows that we have
\begin{moneq}
\deltaLkt(\gt\cdot X)
=
\deltaLkt(X)
\end{moneq}
for all $\gt$ in the connected component of $\MpknR$ containing the identity.
As $\SpknR$ has $2$ connected components (for $k\ge1$) and as $\MpknR$ is a double covering of $\SpknR$, $\MpknR$ has $1$, $2$, or (at most) $4$ connected components.

To see what happens on the other connected components, we first look at the element $\gt_o\in\MpknR$ which is not the identity but whose projection $\projectiongreen\rhoMp(\gt_o)$ in $\SpknR$ is the identity.
According to \recalt{compatibilityMpactionandSpactiononCanLagFr} the action of $\gt_o$ projects to the identity on $\CanLagFrp$.
Hence it can only permute the two elements of the covering map (in a fiber $\CanLagFrpt\to\CanLagFrp$), which means that its left-action coincides with the right-action of the element $(\oneasmat,-1)\in \MlnC$.
The diagonal action of $\gt_o$ thus coincides with the right-action of $\bigl((\oneasmat,-1),(\oneasmat,-1)\bigr)\in \Mlkd$.
But according to our formula, the action of this element changes the value of $\deltaLkt$ with $(-1)^2=1$, \ie, not at all. 
Hence $\deltaLkt$ is invariant under the action of $\gt_o$.

The other connected components are related to the connected component of $\Gl(k,\RR)$ not containing the identity.
A representative of this component is a diagonal matrix $g$ of the form \recalf{defofsubgroupSpk2nR} with $T_{1r}$ and $T_{4r}$ the identity,  $A_g$ diagonal with all diagonal elements except one $1$, the remaining one $-1$, and all other (sub) matrices zero.
We now choose (one out of two) an element $\gt\in \MpknR$ such that $\projectiongreen\rhoMp(\gt)=g\in \SpknR$.
Combining \recalt{formofalphatinMlnC} and \recalf{lemmaonreducedformsandactions}, it follows that we must have
\begin{moneq}
\alphat^{(n)}(\gt,W_i) =
(\ \begin{pmatrix} A_g & *\\ \mathbf0 & \oneasmat \end{pmatrix}
\ ,\ 
z
\ )
\qquad\text{with}\qquad
z^2 = \det(A_g) = -1
\mapob.
\end{moneq}
Using \recalt{compatibilityMpactionandSpactiononCanLagFr} and (again) \recalf{lemmaonreducedformsandactions}, it follows that $\deltaLkt$ changes under the diagonal action of $\gt$ by a factor $\zb\cdot z = 1$.
\end{preuve}

\begin{definition}{Definition}
With these preparations, we can finally define the function $\deltat_D:\PairLagFrpDt[D]M\to \CC$.
The fact that $\PairLagFrpDt[D]M$ is associated to $\SymFrtD M$ gives us trivializations $\Xit_\alpha: \projectiongreen{ \bigl(\pitPairLag\bigr)\mo} (U_\alpha) \to U_\alpha \times \PairCanLagFrpt[n\vert k]$ whose associated transition functions are given by
\begin{moneq}
(\,\Xit_\alpha\scirc \Xit_\beta\mo)(m,X)
=
\bigl( m, \gt\AB(m)\cdot X\bigr)
\mapob,
\end{moneq}
where $\gt\AB\in \MpknR$ are the transition functions of the bundle $\SymFrtD M$.
We now define the continuous functions $\deltat_{D,\alpha}:\projectiongreen{ \bigl(\pitPairLag\bigr)\mo}(U_\alpha) \to \CC$ by
\begin{moneq}
\deltat_{D,\alpha}\bigl(\,  \Xit_\alpha\mo(m,X)\bigr) = \deltaLkt(X)
\mapob,
\end{moneq}
a definition analogous to \recalt{localexpressiondeltasubD}.
As the function $\deltaLkt$ is invariant under the (diagonal) left-action of $\MpknR$, it follows that the locally defined functions $\deltat_{D,\alpha}$ coincide on overlaps and thus define a global continuous function $\deltat_D : \PairLagFrpDt[D]M\to \CC$.

\end{definition}

\begin{proclaim}[thepropertiesofdeltatsubD]{Lemma}
The function $\deltat_D : \PairLagFrpDt[D]M\to \CC$ has the following properties:
\begin{enumerate}
\item
$\bigl( \deltat_D(\ut,\vt)\bigr)^2 = \delta_D\bigl( \projectiongreen\pPairLag(\ut,\vt)\bigr)$;

\item
$\deltat_D\bigl((\ut,\vt)\cdot (\gt_1, \gt_2) \bigr) = \deltat_D(\ut,\vt)\cdot \zb_1\cdot z_2\cdot \vert \det(A)\vert\mo$.

\end{enumerate}

\end{proclaim}

\begin{proclaim}{Corollary}
Let $P_1$ and $P_2$ be two compatible positive Lagrangian distributions and let $\Frt P_1$ and $\Frt P_2$ be the metalinear frame bundles for them induced from the metaplectic frame bundle. Then $\Frt P_1$ and $\Frt P_2$ are compatible.

\end{proclaim}

\begin{preuve}
Let $D$ be the isotropic distribution defined by $D^\CC = \Pb_1\cap P_2$.
Then, according to \recalt{thepropertiesofdeltatsubD} and \recalt{restrictionofdeltasubDgivesdeltak}, the restriction of $\deltat_D:\PairLagFrpDt[D]M\to \CC$ to the subbundle $\PairPolPt{12}\subset \PairLagFrpDt[D]M$ \recalt{PairPolPtsubbundlePairLagFrpDt} has the required properties \recalf{deltaktrootofdeltak} and \recalf{behaviourdeltaktunderMlkd}.
\end{preuve}

\section{Appendix: Fibre bundles and all that}

In this appendix we recall some basic facts about fiber bundles, mainly to introduce notation that is used in the main text. Proofs and details can be found in any textbook on differentiable manifolds that discuss the notion of fiber bundles.

\begin{definition}[definitionofafiberbundle]{Definition}
A (smooth) map $\pi:B\to M$ between manifolds is called a \stresd{(locally trivial) fiber bundle with typical fiber $F$ and structure group $H$} if we can produce the following data:
\begin{enumerate}[(FB1)]
\item\label{conditionFB1}
a collection $\atlas = \{U_\alpha \mid \alpha\in I\}$ of open subsets $U_\alpha\subset M$ covering $M$ (\ie, $\cup_{\alpha\in I} U_\alpha = M$) called a \stresd{trivializing atlas}, the elements of which are called \stresd{(local) trivializing charts},

\item\label{conditionFB2}
for each couple $\alpha,\beta\in I$ a differentiable map $t\AB:U_\alpha \cap U_\beta\to H$ called a \stresd{transition function},

\item\label{conditionFB3}
for each $\alpha\in I$ a diffeomorphism $\Phi_\alpha:\pi\mo(U_\alpha) \to U_\alpha\times F$ and finally

\item\label{conditionFB4}
a left-action of $H$ on $F$.

\end{enumerate}
These data should be compatible in the sense that they should satisfy the conditions
\begin{enumerate}[(FB1)]
\setcounter{enumi}{4}
\item\label{conditionFB5}
on each $\pi\mo(U_\alpha)$ we have  $\pi = \pi_1\scirc \Phi_\alpha$, where $\pi_1:U_\alpha\times F\to U_\alpha$ is the projection on the first coordinate, \ie, we have a commutative diagram
$$
\begin{matrix}
\pi\mo(U_\alpha)  && \rlap{\hss$\overset{\textstyle\Phi_\alpha}{{-}\!{-}\!{-}\!{-}\!{-}\!{-\!}\!\!\to}$} && U_\alpha\times F
\\
\noalign{\vskip1\jot}
& \llap{\raise-5pt\hbox{$\pi$}}  \searrow & & \swarrow \rlap{\raise-5pt\hbox{$\pi_1$}}
\\
&& U_\alpha
\end{matrix}
$$

\item\label{conditionFB6}
for each couple $\alpha,\beta\in I$ the maps $\Phi_\alpha$, $\Phi_\beta$ and $t\AB$ are related by
$$
\Phi_\alpha\scirc \Phi_\beta\mo 
\quad:\quad 
\begin{matrix} 
(U_\alpha\cap U_\beta) \times F 
&\to& 
(U_\alpha\cap U_\beta) \times F
\\
\noalign{\vskip2\jot}
(m,f) 
&\mapsto& 
(m,t\AB(m)\cdot f)
\mapob,
\end{matrix}
$$

\item\label{conditionFB7}
for each triplet $\alpha,\beta,\gamma\in I$ the maps $t\AB$, $t_{\beta\gamma}$ and $t_{\alpha\gamma}$ satisfy the \stresd{cocycle condition}
$$
\forall m\in U_\alpha \cap U_\beta\cap U_\gamma 
\quad:\quad 
t\AB(m) \cdot t_{\beta\gamma}(m) = t_{\alpha\gamma}(m)
\mapob.
$$

\end{enumerate}
Note that the cocycle condition (FB\ref{conditionFB7}) implies that we must have $t_{\alpha\alpha}(m) = e$ the identity element in $H$ (choose $\alpha=\beta=\gamma$) and $t_{\beta\alpha}(m) = t\AB(m)\mo$ (choose $\gamma=\alpha$).

A \stresd{principal fiber bundle with structure group $G$} is a fiber bunde $\pi:B\to M$ with typical fiber $F=G$ and structure group $H=G$ such that the action of the structure group $H=G$ on the typical fiber $F=G$ is just left-translation. If that is the case, there is a natural right-action of $G$ on $B$ which is compatible with the trivializations $\Phi_\alpha$ in the sense that we have
$$
\Phi_\alpha(b) = (m,g)
\quad\Rightarrow\quad
\Phi_\alpha(b\cdot h) = (m,gh)
\mapob.
\leqno \text{(FB8)}
$$
Moreover, this right-$G$-action on $B$ is free and the quotient (orbit) space $B/G$ is the base manifold $M$.

\end{definition}

\begin{definition}{Remarks}
\myitemize
Condition (FB\ref{conditionFB5}) implies that the map $\Phi_\alpha \scirc \Phi_\beta\mo$ must be of the form $(m,f) \mapsto (m,\Phi\AB(m,f)$ with the map $f\mapsto \Phi\AB(m,f)$ a diffeopmorphism of $F$ for fixed $m\in U_\alpha\cap U_\beta$. 
Condition (FB\ref{conditionFB6}) requires that these diffeomorphisms belong to a specified structure group. And it is the presence of the structure group that provides us with different kinds of fiber bundles. 
For instance, if $F$ is a vector space and $H$ is (a subgroup of) the group of linear isomorphisms of $F$, then one speaks of a vector bundle. And if the structure group is $F$ acting by translations, one speaks of a principal fiber bundle.

\myitemize
On $(U_\alpha\cap U_\beta\cap U_\gamma)\times F$ we obviously have
$$
(\Phi_\alpha\scirc \Phi_\beta\mo) \scirc (\Phi_\beta \scirc \Phi_\gamma\mo) = \Phi_\alpha\scirc \Phi_\gamma
$$
and thus by (FB\ref{conditionFB6}) we must have
$$
t\AB(m)\cdot t_{\beta\gamma}(m)\cdot f = t_{\alpha\gamma}(m)\cdot f
$$
for all $f\in F$. If the action of $H$ on $F$ is free then we get (FB\ref{conditionFB7}) automatically. Only if it is not, we need to add it.

\myitemize
The standard definition of a principal fiber bundle with structure group $G$ is a smooth map $\pi:B\to M$ and a (smooth) right-action of $G$ on $B$ satisfying the conditions (FB\ref{conditionFB1}), (FB\ref{conditionFB3}) and (FB\ref{conditionFB5}) with $F$ replaced by $G$ as well as the additional condition that the right-action of $G$ on $B$ must be compatible with the standard right-action of $G$ on itself via the maps $\Phi_\alpha$ in the sense that we must have
$$
\Phi_\alpha(b) = (m,h) \quad \Rightarrow\quad 
\Phi_\alpha(b\cdot g) = (m,h\cdot g)
\mapob.
\leqno \text{(FB8)}
$$
It follows immediately that the $G$-action on $B$ must be free. And if we look at the maps $\Phi_\alpha\scirc \Phi_\beta\mo$, it follows that we must have the implication
$$
(\Phi_\alpha\scirc \Phi_\beta\mo)(m,h) = (m,k)
\qquad\Longrightarrow\qquad
(\Phi_\alpha\scirc \Phi_\beta\mo)(m,hg) = (m,kg)
\mapob.
$$
If we define $t\AB(m)$ via $(\Phi_\alpha\scirc \Phi_\beta\mo)(m,e) = (m,t\AB(m))$, we get (FB\ref{conditionFB6}) with indeed left-translation of $G$ on itself:
$$
(\Phi_\alpha\scirc \Phi_\beta\mo)(m,g) = (m,t\AB(m)\cdot g)
$$
We thus have recovered (FB\ref{conditionFB2}), (FB\ref{conditionFB4}) (in the form of left-translation) and (FB\ref{conditionFB6}). And since left-translation of $G$ on itself is a free action, we also have (FB\ref{conditionFB7}).

\myitemize
Any local trivialization of a principal fiber bundle $\pi:B\to M$ with structure group $G$ determines a local section, but more importantly, any local section determines a local trivialization. 
More precisely, if $\Phi_\alpha:\pi\mo(U_\alpha)\to U_\alpha\times G$ is a local trivialization, we obtain a local section $s_\alpha:U_\alpha\to \pi\mo(U_\alpha)$ by
\begin{moneq}[sectionassociatedtotrivialization]
s_\alpha(m) = \Phi_\alpha\mo(m,e)
\mapob.
\end{moneq}
Conversely, if $s_\alpha:U_\alpha\to \pi\mo(U_\alpha)$ is a local (smooth) section, the map $\Psi_\alpha:U_\alpha\times G\to \pi\mo(U_\alpha)$ defined by
$$
\Psi_\alpha(m,g) = s_\alpha(m)\cdot g
$$
is a diffeomorphism whose inverse $\Phi_\alpha=\Psi_\alpha\mo$ is a local trivialization such that we have \recalf{sectionassociatedtotrivialization}.
If $s_\beta:U_\beta\to \pi\mo(U_\beta)$ is another local (smooth) section, there exists a (smooth) function $g\AB:U_\alpha\cap U_\beta\to G$ such that for $m\in U_\alpha\cap U_\beta$ we have
$$
s_\beta(m) = s_\alpha(m)\cdot g\AB(m)
\mapob.
$$
It immediately follows that the local trivializations $\Phi_\alpha$ and $\Phi_\beta$ are related by
$$
(\Phi_\alpha\scirc \Phi_\beta\mo)(m,g) = \bigl(m,g\AB(m)\cdot g\bigr)
\mapob.
$$

\end{definition}

\begin{definition}[reconstructionofafiberbundle]{Reconstruction of a fiber bundle}
Suppose we have a manifold $M$, a collection $\atlas=\{\, U_\alpha\mid \alpha\in I \,\}$ of open subsets $U_i\subset M$ covering $M$ and a collection of functions $t\AB:U_\alpha \cap U_\beta \to H$ with values in a Lie group $H$ satisfying the cocycle condition (FB\ref{conditionFB7}). In addition, the Lie group $H$ is suppsed to act smoothly on the left on a manifold $F$. Then we can (re)construct a fiber bundle $\pi:B\to M$ with typical fiber $F$ and structure group $H$ such that $\atlas$ is a trivializing atlas and the functions $t\AB$ the corresponding transition functions.

The construction starts by considering the disjoint union $\widetilde B = \smash{\coprod\limits_{\alpha\in I}} (U_\alpha \times F)$ and to define an equivalence relation $\sim$ on $\widetilde B$ by
$$
(m_\alpha,f_\alpha) \in U_\alpha \times F \ \sim \ (m_\beta,f_\beta) 
\qquad\Longleftrightarrow\qquad
m_\alpha=m_\beta \text{ and } f_\alpha = t\AB(m_\beta)\cdot f_\beta
\mapob.
$$
The manifold $B$ then is defined as the set of equivalence classes with respect to this equivalence relation:
$$
B=\widetilde B/\sim
\mapob.
$$
The projection $\pit:\widetilde B\to M$ defined by $\pit(m,f) = m$ is compatible with the equivalence relation and thus induces a projection $\pi:B\to M$. 
It then is straightforward to check that $\pi:B\to M$ is a fiber bundle with typical fiber $F$ and structure group $H$ for which $\atlas$ is a trivializing atlas and the $t\AB$ the corresponding transition functions.

\end{definition}

\begin{definition}[constructionofassociatedbundles]{Construction of associated bundles}
Let $\pi_P:P\to M$ be a principal fiber bundle with structure group $G$ (acting on the right on $P$), let $\rho:G\to H$ be a homomorphism of Lie groups and let $H$ act on a manifold $F$. With these ingredients we can construct a fibre bundle $\pi_B:B\to M$ with typical fiber $F$ as follows. One considers the manifold $P\times F$ on which we let the group $G$ act on the right by
$$
(p,f)\cdot g = (p\cdot g, \rho(g\mo)\cdot f)
\mapob.
$$
The manifold $B$ is defined as the quotient (orbit) space $B=(P\times F)/G$ and we will denote the canonical projection by $\pi_\sim:P\times F\to B$. For this quotient space one also finds the following notation:
$$
B = P\times_{G,\rho} F
\mapob.
$$
To define the projection $\pi_B$ we note that we have
\begin{align*}
(\pi_P \scirc \pi_1)\bigl( (p,f)\cdot g\bigr) 
&= 
(\pi_P \scirc \pi_1)\bigl( (p\cdot g, \rho(g\mo)\cdot f) \bigr)
=
\pi_P(p\cdot g)
=
\pi_P(p)
\\&
= 
(\pi_P\scirc \pi_1)\bigl( (p,f) \bigr)
\mapob,
\end{align*}
\ie, the map $\pi_P\scirc \pi_1$ is constant on the $G$-orbits, where $\pi_1$ denotes the projection on the first factor. It follows that there exists a unique map $\pi_B:B\to M$ such that $\pi_B\scirc \pi_\sim = \pi_P\scirc \pi_1$.

In terms of transition functions this construction is fairly easy to understand. Let $\atlas$ be a trivializing atlas for the principal fiber bundle $\pi_P:P\to M$ with associated transition functions $g\AB:U_\alpha\cap U_\beta\to G$. Then the bundle $\pi_B:B\to M$ is descibed by $\atlas$ and the transition functions $t\AB=\rho(g\AB)$ according to the construction \recalt{reconstructionofafiberbundle}.
A slightly more precise way to say the same is to start with trivializing sections $s_\alpha:U_\alpha\to \pi_P\mo(U_\alpha)$ of $P$. 
Each $s_\alpha$ determines a trivialization $\Phi_\alpha:\pi_B\mo(U_\alpha)\to U_\alpha\times F$ by the formula
$$
\Phi_\alpha\mo(m,f)
=
\pi_\sim\bigl(s_\alpha(m),f\bigr)
\mapob.
$$
As on $U_\alpha\cap U_\beta$ we have $s_\beta(m) = s_\alpha(m) \cdot g\AB(m)$, we thus have
\begin{align*}
(\Phi_\alpha\scirc \Phi_\beta\mo)(m,f)
&
=
(\Phi_\alpha\scirc\pi_\sim)\bigl(s_\beta(m),f\bigr)
\\&
=
(\Phi_\alpha\scirc\pi_\sim)\bigl(s_\beta(m)\cdot g_{\beta\alpha}(m),\rho\bigl(g\AB(m)\bigr)\cdot f\bigr)
\\&
=
(\Phi_\alpha\scirc\pi_\sim)\bigl(s_\alpha(m),\rho\bigl(g\AB(m)\bigr)\cdot f\bigr)
=
\bigl(m,\rho\bigl(g\AB(m)\bigr)\cdot f\bigr)
\mapob,
\end{align*}
showing that the transition functions for the associated bundle are indeed $\rho(g\AB)$.

\end{definition}

\begin{definition}{Remarks}
\myitemize
If $\pi_B:B\to M$ is a fiber bundle with typical fiber $F$ and structure group $H$, we can choose a trivializing atlas $\atlas$ with associated transition functions $t\AB:U_\alpha\cap U_\beta\to H$. 
And if $\rho:H\to H'$ is a Lie group homomorphism and if the Lie group $H'$ acts on a manifold $F'$, we can contruct a fiber bundle $\pi_{B'}:B'\to M$ using the construction \recalt{reconstructionofafiberbundle} with the data $\atlas$ and functions $t'\AB = \rho(t\AB)$. 
This is exactly what we described in the construction of an associated bundle. However, in general there is no intrinsic way to describe this bundle $B'$ in terms of the initial bundle $B$. 
Such a description is reserved to the situation when we start with a pincipal fiber bundle

\myitemize
On the other hand, for some kind of bundles one can (re)construct an associated principal fiber bundle $P$ from a given fiber bunde $B$. One important case is when we start with a vector bundle, \ie, a fiber bundle with typical fiber a vector space and with structure group (a subgroup of) the group of linear automorphisms. More precisely, if $\pi_B:B\to M$ is a vector bundle, we can consider the bundle $\pi_P:P\to M$ in which the fiber $\pi_P\mo(m)$ consists of all bases of the vector space $\pi_B\mo(m)$. This bundle $P$ is called the frame bundle associated to the vector bundle $B$; it is a principal fiber bundle with structure group $GL(n)$ if the typical fiber of $B$ is of dimension $n$.

\end{definition}

\section{The different ``projections''}
\label{sectionofallprojections}
\noindent Group homomorphisms:
\begin{moneq}
\rhoMl:\MlnC\to \GlnC
\qquad,\qquad
\rhoMp: \MpnR\to \SpnR
\mapob.
\end{moneq}
Compatible polarizations:
\begin{moneq}
\piPairPol:\PairPolP{12} \to M
\quad,\quad
\pitPairPol:\PairPolPt{12}\to M
\quad,\quad
\pPairPol:\PairPolPt{12} \to \PairPolP{12}
\mapob.
\end{moneq}
Symplectic frame bundles:
\begin{moneq}
\piSymp:\SymFr M \to M
\quad,\quad
\pitSymp:\SymFrt M\to M
\quad,\quad
\pSymp:\SymFrt M \to \SymFr M
\mapob.
\end{moneq}
Lagrangian frame bundles:
\begin{moneq}
\piLag:\LagFr M \to M
\quad,\quad
\pitLag:\LagFrpt M\to M
\quad,\quad
\pLagpd:\LagFrpt M \to \LagFrp M
\mapob.
\end{moneq}
Compatible Lagrangian frame bundles:
\begin{moneq}
\piPairLag:\PairLagFrD[D]M \to M
\quad,\quad
\pitPairLag:\PairLagFrpDt[D]M\to M
\quad,\quad
\pPairLag:\PairLagFrpDt[D]M \to \PairLagFrpD[D]M
\mapob.
\end{moneq}
Typical fibres:
\begin{moneq}
\pCanLag:\CanLagFrpt\to\CanLagFrp
\qquad,\qquad
\pPairCanLag:\PairCanLagFrpt[n\vert k]\to\PairCanLagFrp[n\vert k]
\end{moneq}
The leaf space:
\begin{moneq}
\pileaf:M\to M/D
\end{moneq}

\section*{Acknowledgements}

This work was supported in part by the Labex CEMPI  (ANR-11-LABX-0007-01).

\bibliographystyle{amsalpha}

\providecommand{\bysame}{\leavevmode\hbox to3em{\hrulefill}\thinspace}
\providecommand{\MR}{\relax\ifhmode\unskip\space\fi MR }
\providecommand{\MRhref}[2]{%
  \href{http://www.ams.org/mathscinet-getitem?mr=#1}{#2}
}
\providecommand{\href}[2]{#2}

\end{document}